\documentclass[10pt, conference, compsocconf]{IEEEtran} %

\usepackage{booktabs}
\usepackage{graphicx}
\usepackage{cite}
\newcommand{\ignore}[1]{}
\usepackage{fancyhdr}
\usepackage[normalem]{ulem}
\usepackage[hyphens]{url}
\usepackage[hidelinks]{hyperref}

\usepackage{multirow}
\let\labelindent\relax
\usepackage{enumitem}
\usepackage[usenames,dvipsnames]{xcolor}
\usepackage{amsmath}

\usepackage[english]{babel}
\usepackage{blindtext,tikz}
\usetikzlibrary{calc}

\usepackage{subcaption}
\DeclareCaptionLabelSeparator{periodspace}{.\quad}
\begin{document}

\title{Using ECC DRAM to Adaptively Increase Memory Capacity\vspace{-10mm}}
\author{\large
  Yixin Luo$^\dagger$ \quad
  Saugata Ghose$^\dagger$ \quad
  Tianshi Li$^\dagger$ \quad
  Sriram Govindan$^\S$ \\
  Bikash Sharma$^\S$ \quad
  Bryan Kelly$^\S$ \quad
  Amirali Boroumand$^\dagger$ \quad
  Onur Mutlu$^{*,\dagger}$ \vspace{2mm} \\

    $^\dagger$Carnegie Mellon University \quad
    $^\S$Microsoft Corporation \quad
    $^*$ETH Z{\"u}rich
}

\maketitle
\pagestyle{plain}



\begin{abstract}

Modern DRAM modules are often equipped with hardware error correction
capabilities, especially for DRAM deployed in large-scale data centers, as
process technology scaling has increased the susceptibility of these devices to
errors.  To provide fast error detection and correction, error-correcting codes
(ECC) are placed on an additional DRAM chip in a DRAM module.  This
additional chip expands the raw capacity of a DRAM module by 12.5\%, but the
applications are unable to use any of this extra capacity, as it is used
exclusively to provide reliability for \emph{all} data.  In reality, there are a
number of applications that do not need such strong reliability for all their
data regions (e.g., some user batch jobs executing on a public
cloud), and can instead
benefit from using additional DRAM capacity to store extra data.  Our goal in this work
is to provide the additional capacity within an ECC DRAM module to applications
when they do \emph{not} need the high reliability of error correction.



In this paper, we propose \emph{Capacity- and Reliability-Adaptive Memory}
(CREAM), a hardware mechanism that adapts error-correcting DRAM modules to 
offer multiple levels of error protection, and provides the capacity saved from
using weaker protection to applications.  For regions of memory
that do not require strong error correction, we either provide no ECC protection
at all, or provide error detection in the form of multi-bit parity.  We evaluate
several layouts for arranging the data within ECC DRAM in these reduced-protection modes, taking
into account the various trade-offs exposed from exploiting the extra chip.  Our
experiments show that the increased capacity provided by CREAM improves
performance by 23.0\% for a memory caching workload for databases, and by 37.3\% for a
commercial web search workload executing production query traces.  In addition,
CREAM can increase bank-level parallelism within DRAM, offering further 
performance improvements.

\end{abstract}


\section{Introduction}
\label{sec:introduction}


Error-correcting DRAM modules are heavily used in servers and data centers today, as DRAM
has become increasingly susceptible to errors due to continued process
technology scaling~\cite{schroeder2009dram, hwang2012cosmic, sridharan2012study,
sridharan2013feng, sridharan2015memory, nightingale2011cycles, 
li2010realistic, meza2015revisiting, kim.isca14}.  By storing \emph{error-correcting codes} 
(ECC) within error-correcting DRAM modules, error detection and correction is
performed in hardware.  Today, most error-correcting (or ECC) DRAM modules employ single error correction, 
double error detection (SECDED) codes~\cite{hsiao1970class}.

Error correction is performed when a memory request reads or writes data.
For widely-used DDR3 and DDR4 DRAM, these requests are performed 64~bytes at a time.  In order to limit the width of the
off-chip bus between the processor and the DRAM module, this data is sent in several 
smaller \emph{data bursts} (e.g., eight 64-bit data bursts for DDR3
and DDR4 DRAM).  For every 64-bit data burst, an 8-bit SECDED code is 
transmitted alongside the data to the \emph{memory controller}, which interfaces
between the processor and the DRAM module.  For each burst, the 8-bit SECDED code is used
to determine if an error exists in the 64-bit burst, and if so, an error
correction algorithm is applied within the controller to correct the data.  In
all, for the eight bursts of data sent, an ECC DRAM module contains 8~bytes worth of
correction information.  On the module, this correction data is stored on an
\emph{additional DRAM chip}, which operates in lockstep with the DRAM chips on
the module that contain the data, and provides error correction for \emph{all}
of the data in memory.

An ECC DRAM provides \emph{reliability} at the expense of additional memory
\emph{capacity}. The key question we ask in this study is: \emph{Can we use the
additional capacity of the extra chip in ECC DRAM when memory regions of applications do not need
the reliability it provides?} We make two \emph{key observations} about the 
trade-off between reliability and capacity.

First, there are many applications that benefit from additional DRAM capacity.
\emph{Page faults} are costly operations, taking hundreds of microseconds to
retrieve data not mapped in DRAM.
Several works have demonstrated that with additional DRAM capacity, application
performance improves significantly, as the additional capacity helps to significantly
reduce the number of page faults that take place~\cite{lim2009disaggregated,
anagnostopoulou2012barely, zhou2004dynamic, chu1972page, gupta1978working}.
We confirm this behavior when we analyze data-intensive server workloads, which include a
commercial web search application from Microsoft's production data centers

Second, there are many instances where workloads or memory regions may not benefit from
error correction.  This primarily happens for two reasons:  (1)~Several applications 
are resilient to errors, or are of low importance to server owners, and 
therefore do not require full error correction~\cite{luo.dsn14, 
song.asplos11, sampson11, approx12, baek.pldi10}.  For example, for WebSearch, a very small
number of incorrect query responses does not significantly affect user quality of
service~\cite{baek.pldi10}.  Likewise, a cloud service provider may have little need to ensure that 
client virtual machines (VMs) operate reliably, and could offer reliability-free
VMs at a lower price to fit a greater number of VMs into each machine for
greater revenue.  (2)~Certain regions of memory may not require full error
correction.  At the hardware level, newer DRAM may be less susceptible to faults, and due to process
variation, there may be regions of DRAM that have very low error
rates~\cite{khan.sigmetrics14, schroeder2009dram, hwang2012cosmic,
sridharan2012study}. At the software level, some
data regions of an application may not need any correction as well~\cite{luo.dsn14}.

\begin{figure}[h!]
\centering
\vspace{-7pt}
\includegraphics[trim=0 105 0 0,clip,width=0.9\linewidth]{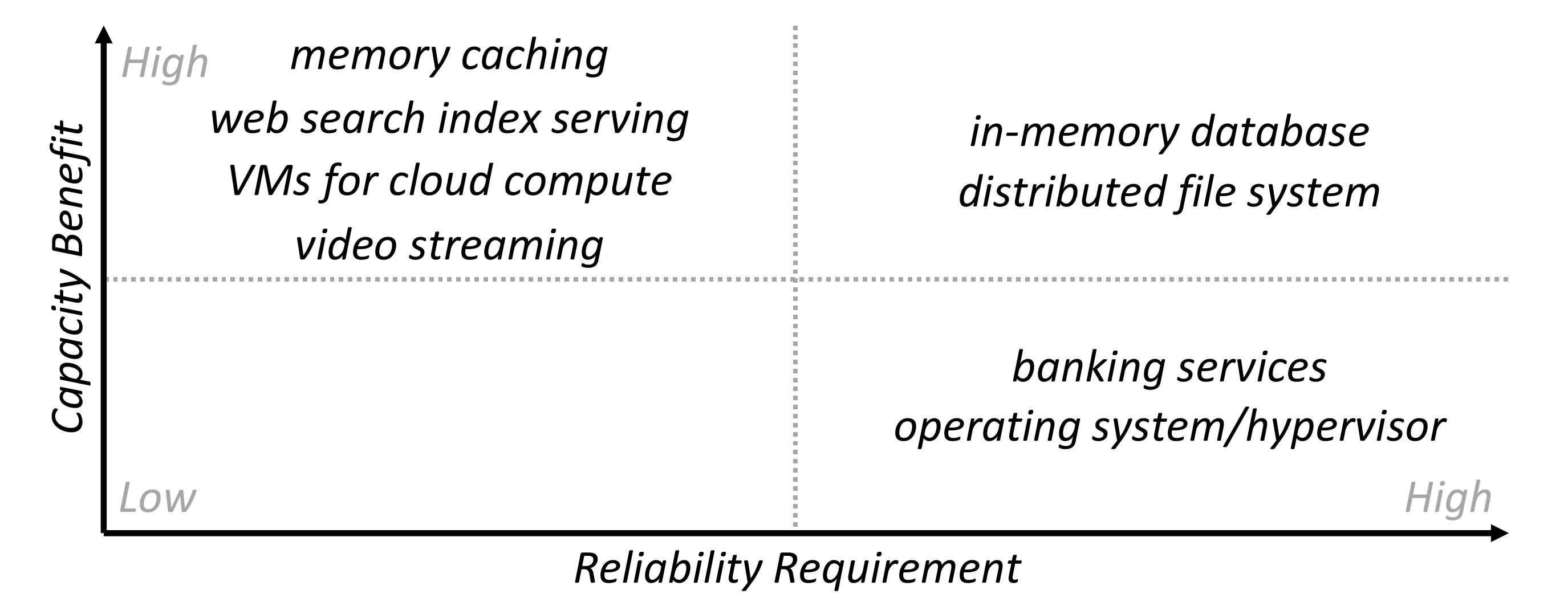}%
\vspace{-5pt}
\caption{
  Memory reliability requirements and memory capacity benefits for example applications.
}
\label{fig:introduction}
\vspace{-5pt}
\end{figure}

\emph{Our goal} in this work is to \emph{enable} the additional capacity within
ECC DRAM modules for applications \emph{when SECDED reliability is not  
required} during their execution, while continuing to provide error correction
for applications that need reliability.  Figure~\ref{fig:introduction} shows the space of 
applications across the dimensions of reliability and capacity, and shows several 
example applications within each quadrant of the space.
For applications (or memory regions) that require high reliability, but do not benefit from
additional data capacity, ECC should continue to work as it
has in the past, providing quick hardware error correction.  For applications
that do not require high reliability, but benefit from additional data capacity, we aim to
convert the space used by ECC data in DRAM into additional data capacity.  For those
applications that require reliability \emph{and} benefit from capacity, we
aim to support a lower-strength reliability mechanism that allows for
\emph{some, but not all} of the ECC capacity to be converted into additional
data capacity. At a finer granularity, the reliability requirements of memory
regions also vary~\cite{luo.dsn14}.

To this end, we propose \emph{Capacity- and Reliability-Adaptive Memory} (CREAM), a new 
hardware mechanism that take advantages of the additional DRAM capacity that currently
goes underutilized for applications (or memory regions) that do not require high reliability.  CREAM
provides two capabilities.  First, it converts a portion of the space in an ECC DRAM
module into \emph{non-ECC mode}, freeing up the space in the additional ECC chip
so it can store application data.  We propose three solutions that expose all of
this capacity to applications: (1)~a method that requires no changes to the ECC 
DRAM module, using additional reads and writes issued by the memory controller
to access the extra space; (2)~a method that adds simple logic to the DRAM 
module to reduce the write overhead to the extra space, and (3)~a method that
reorganizes the entire data layout so that instead of accessing nine chips at a
time in each of the eight banks, we can access only eight chips at a
time, allowing us to use the leftover chips as an additional DRAM bank. Second,
CREAM converts part of the space in an ECC DRAM module into \emph{parity
mode}, where parity checks are provided instead of full-blown SECDED correction,
allowing applications to maintain lower-strength reliability while still 
benefiting from additional data space.

We perform two studies to gauge the effectiveness of CREAM.  First, we evaluate
CREAM on large-memory workloads.  We execute \emph{production query traces} on
a commercial web search application from Microsoft, and find 
that the 12.5\% increase in DRAM capacity provided by CREAM improves the workload's overall system
performance by 37.3\%.  We also find that CREAM improves the performance of
a \emph{memcached} database workload by 23.0\%, including all overheads.  Second, we find
that that the increased bank-level parallelism allows CREAM to provide performance gains (0.8\%
for \texttt{memcached}, and 2.4\% on average across 40 multiprogrammed workloads), 
\emph{on top of} the gains from having additional effective memory capacity.

In this work, we make the following contributions:

\begin{itemize}[leftmargin=*,noitemsep,topsep=0pt]

  \item We provide a simple and practical mechanism 
    to efficiently harness part or all of the additional space previously set aside for
    error correction within an ECC DRAM module, providing additional data
    capacity to applications and memory regions that don't require high reliability.

  \item We propose three methods of increasing data capacity by 12.5\% when applications
    or memory regions do not require error correction or detection.  One of these methods 
    increases both DRAM capacity \emph{and} bank-level parallelism, providing 
    additive performance improvements.

  \item We propose a method of exposing additional data capacity without
    fully eliminating reliability, by supporting multi-bit parity for lightweight
    error detection.  We evaluate this method quantitatively.

  \item Our evaluations with major data-intensive applications show that using
    the additional space that is otherwise dedicated for ECC improves
    performance significantly, mitigating the high penalty of page faults.

\end{itemize}


\section{Background}
\label{sec:background}

To understand the opportunities available for expanding memory capacity when
strong reliability is not required, we first provide necessary background on
DRAM organization and error correction.


\subsection{DRAM Organization}
\label{sec:background:organization}

DRAM communicates with the processor across a DRAM \emph{channel},
an \emph{off-chip} bus used to send DRAM commands and data.  
For DDR3 and DDR4 
DRAM, this channel is only 64~\emph{bits} wide, and is used to communicate a single
piece of data at a time (known as a \emph{data burst}).  DRAM performs
operations at the granularity of a \emph{64-byte} \emph{cache line}.
As a result, \emph{eight} back-to-back data bursts are
required to send a single cache line of data.  Data requests are managed by a
\emph{memory controller}, which typically resides on-chip with the processor.
The memory controller receives per-cache-line memory requests, and breaks these
requests down into a series of DRAM commands that are issued to DRAM.

\begin{figure}[h!]
  \centering
  \vspace{-10pt}
  \begin{subfigure}[b]{0.38\linewidth}
    \centering
    \includegraphics[trim = 250 115 340 65, clip, width=.85\textwidth]{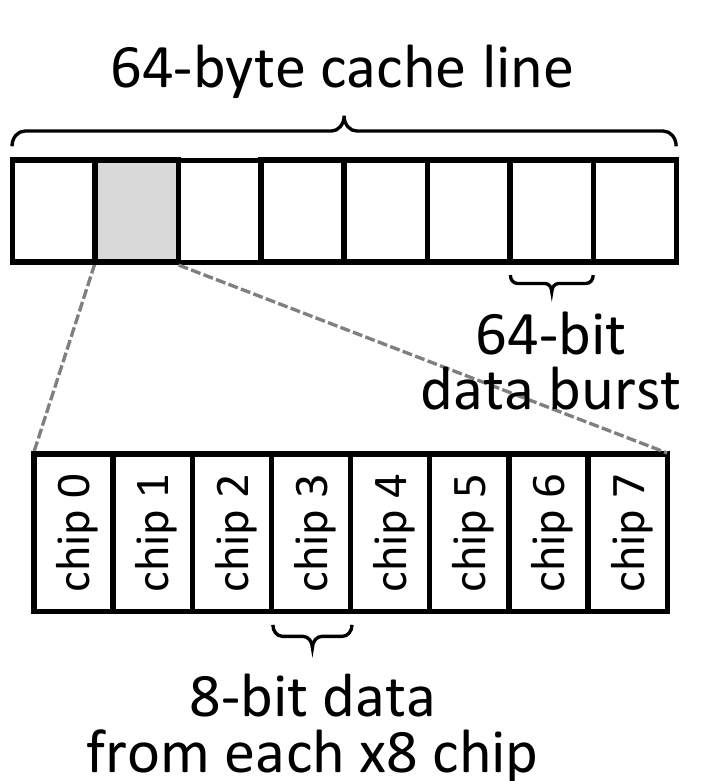}\vspace{-2pt}%
    \caption{\scriptsize Cache line breakdown.}
    \label{fig:cache_line}
  \end{subfigure}%
  \qquad%
  \begin{subfigure}[b]{0.53\linewidth}
    \includegraphics[trim = 75 118 375 65, clip, width=0.95\textwidth]{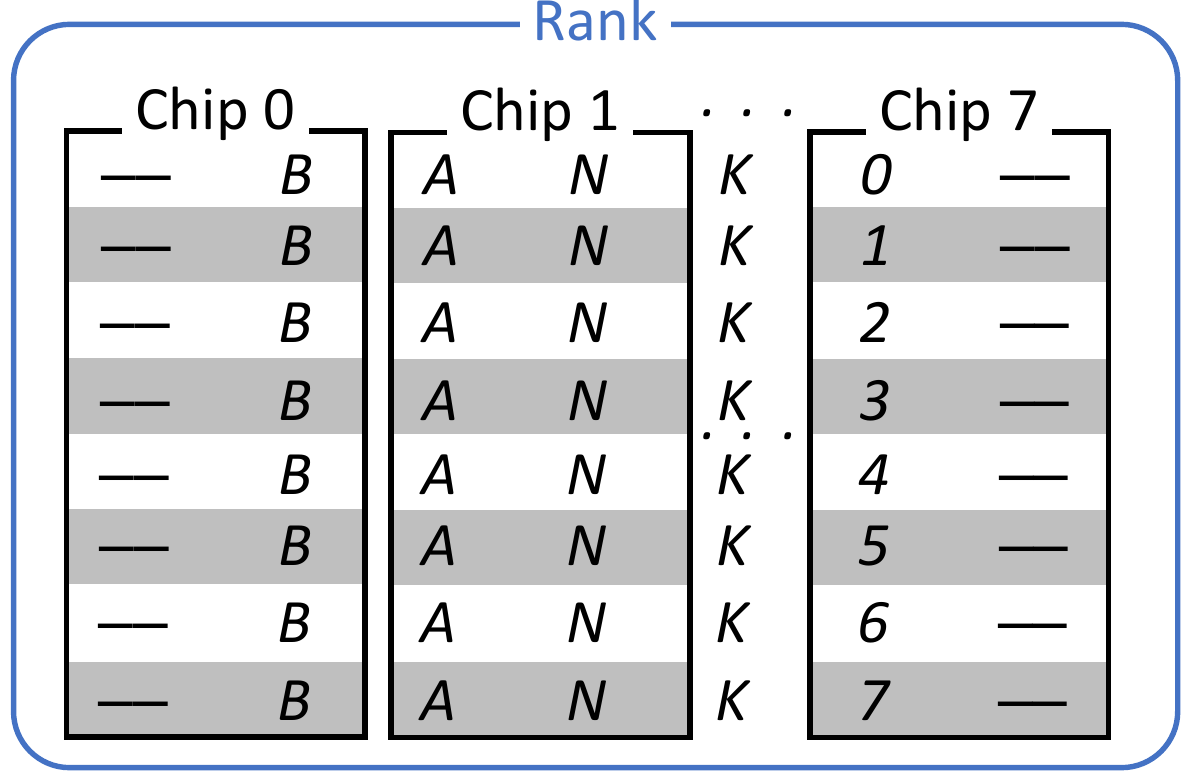}\vspace{-2pt}%
    \caption{\scriptsize Chip arrangement within a rank.}
    \label{fig:rank_layout}
  \end{subfigure}%
  \vspace{-5pt}
  \caption{DRAM organization with x8 chips.}%
  \label{fig:organization}
  \vspace{-18pt}
\end{figure}

A DRAM module (i.e., a \emph{DIMM}, or dual inline memory module) is made up of
several DRAM \emph{chips}.  Each chip has a fixed \emph{data width} (i.e., the
amount of data that it can transmit at any given time).  For example, an x8 DRAM
chip can transmit 8~bits of data at a time.
Several of these chips work in \emph{lockstep} to provide 64~bits of
partial data from a single cache line, 
as shown in 
Figure~\ref{fig:cache_line}.  The chips working together in
lockstep are known as a \emph{rank}.  For x8 DRAM chips, each rank contains
eight chips, as shown in Figure~\ref{fig:rank_layout}.  
In order to work in lockstep, the chips within a
rank
share the command and address wires,
ensuring that they all perform the same operation on the same location.

Within each DRAM chip, data is stored within two-dimensional arrays of
capacitive DRAM cells.  
The array is accessed one row at a time, and the row being operated on must be \emph{activated} (i.e. opened), 
which brings the contents of the \emph{entire} row into a \emph{row buffer}.  A memory
request to a row already opened within the row buffer is known as a \emph{row
buffer hit}.  In contrast, if a memory request wants to access a row other than
the one currently open, it must first close the current row (\emph{precharge}),
and then activate the desired row; this is known as a \emph{row buffer miss}.


To increase the probability of a row buffer hit, data is mapped
into the DRAM module to maximize \emph{row buffer locality}, by ensuring that adjacent
columns of data within the same row map to adjacent data within the same OS
page.\footnote{Each row typically contains multiple OS pages, but to simplify
our explanations without loss of generality, we assume throughout this paper 
that each DRAM row contains only a single page.}
In part to increase row buffer locality, the
two-dimensional cell array is split into multiple \emph{banks}, each with its own
row buffer.  These banks can independently service requests in parallel (known
as \emph{bank-level parallelism}).
In DDR3 DRAM, there are
eight banks per chip, and since the chips within a rank operate in lockstep, 
there are effectively eight banks available in each rank (see 
Figure~\ref{fig:rank_layout}).  DDR4 DRAM chips contain 16~banks per rank.


\subsection{Error Protection in Memory}
\label{sec:background:ecc}

Occasionally, DRAM is susceptible to bit errors when data is being read or
written~\cite{kim.isca14,schroeder2009dram, meza2015revisiting}.  These errors can either be \emph{hard}
(i.e., an intrinsic defect within the DRAM itself) or \emph{soft} (i.e., a
transient error that can occur due to phenomena such as cosmic 
rays)~\cite{schroeder2009dram,
hwang2012cosmic, sridharan2012study, sridharan2013feng, sridharan2015memory,
nightingale2011cycles, li2007memory, li2010realistic, meza2015revisiting}.
Memory errors have the potential to greatly impact application stability.  If a
memory error goes undetected, it can lead to \emph{silent data corruption},
and can alter critical data or cause a system crash.
\label{sec:background:memory}

To mitigate these memory errors, a popular DRAM error correction mechanism,
SECDED (single error correction, double error detection) is widely used in
today's server memory~\cite{hsiao1970class}. SECDED can correct one error and detect two
errors, using 8~bits of ECC information for every 64~bits of data, with low
logical complexity.  
A common variant of DRAM that directly encodes SECDED in hardware is known as
\emph{ECC memory}, where \emph{all} of the data within DRAM is protected.  This
allows error protection to be \emph{performed entirely in hardware} as part of
every memory request.  
For every 64-bit data burst during a request, an 8~bit SECDED code (stored in an
additional DRAM chip) is also read out in
lockstep, and transmitted back to the memory controller.  Note that this
expands the off-chip data bus to 72~bits.  Within the
controller, each data burst is checked using the SECDED code to detect whether
an error has occurred, and either correct the data if it can or notify the system
that data has been corrupted.

\begin{figure}[h]
\centering
\vspace{-10pt}
\includegraphics[trim=10 100 10 0,clip,width=\linewidth]{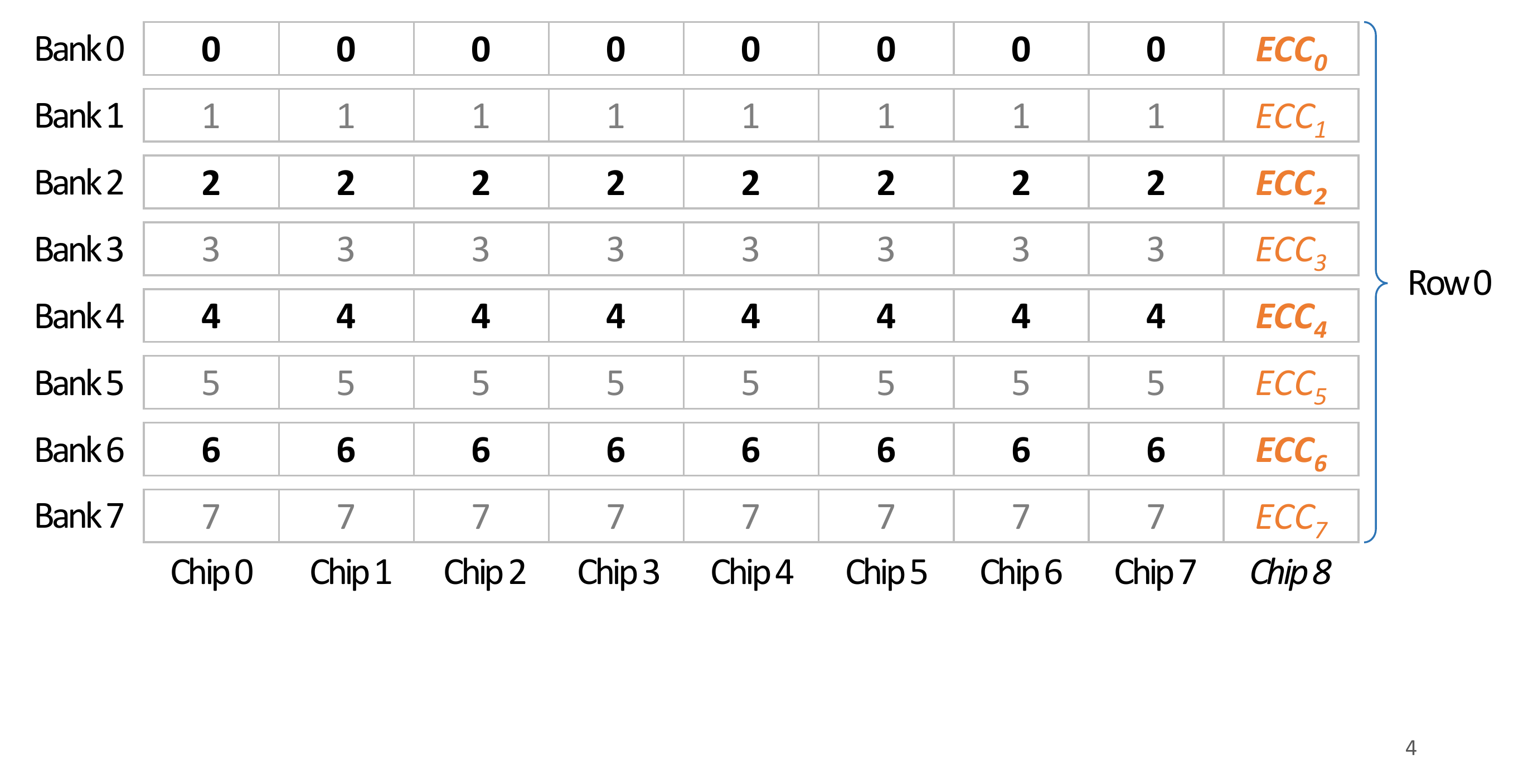}%
\tiny\newline
\newline
\includegraphics[trim=10 150 10 0,clip,width=\linewidth]{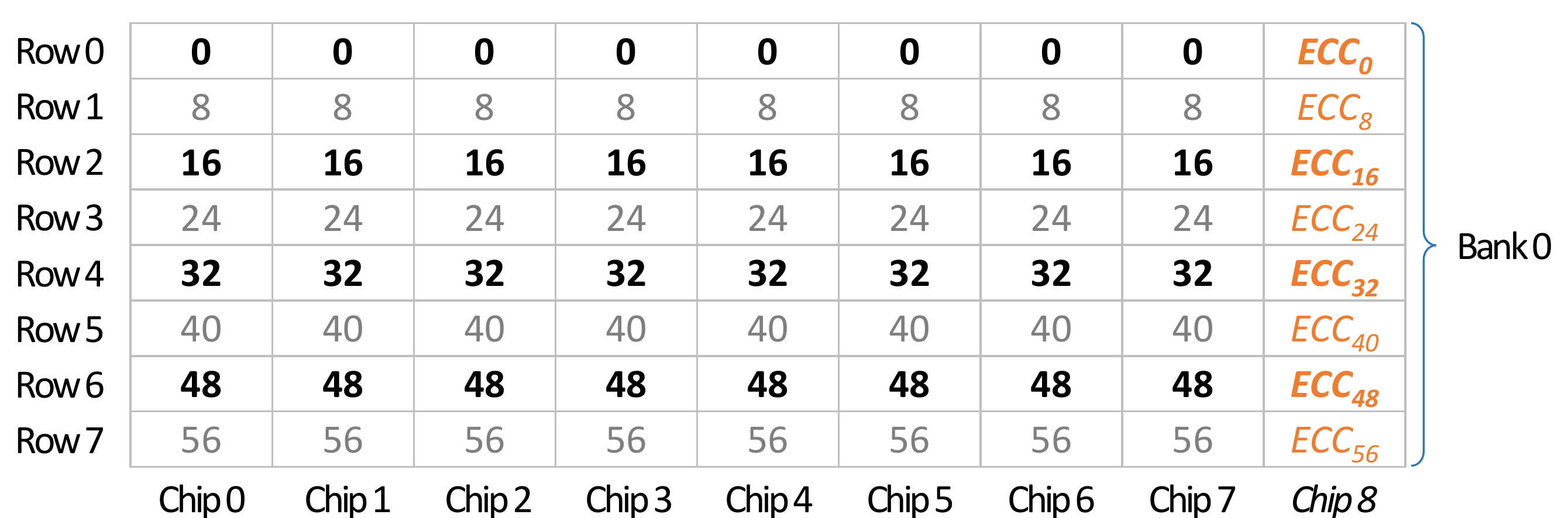}%
\vspace{-5pt}
\caption{
  Data layout of physical pages and ECC information within baseline ECC memory,
  shown for the first row within all eight banks (top) and within the first 
  eight rows of Bank~0 (bottom).
}
\label{fig:baseline}
\vspace{-5pt}
\end{figure}

Figure~\ref{fig:baseline} shows how data pages and ECC are laid out within an
ECC DRAM module.  To simplify our explanations, the data layout figures
in this paper assume that (1)~each DRAM row stores a single OS page, 
(2)~there is a single DRAM channel, and (3)~the DRAM channel contains only a
single rank.  In order to maximize row buffer locality (see 
Section~\ref{sec:background:organization}), we arrange physical pages such that
consecutively-numbered pages map to different banks.
We show the data
layout from two views: the first row across all banks (the top of 
Figure~\ref{fig:baseline}), and the first eight rows within Bank~0 (bottom).

As mentioned above, providing SECDED codes for \emph{all} of the data in DRAM
requires manufactures to add additional chips onto each DRAM module.  The
additional chip expands the \emph{raw capacity} of the DRAM module by 12.5\% 
(since we add 8~bits for every 64~bits of data).  However, the \emph{effective}
DRAM capacity remains unchanged with respect to a DRAM module without ECC 
support, as this additional chip is \emph{exclusively} used to store the 
error-correcting codes.


\section{Motivation: Capacity vs.\ Reliability}
\label{sec:motivation}

DRAM reliability currently takes a one-size-fits-all approach,
providing strong error correction for all data, but this results in significant
reliability over-provisioning, which impacts the revenue of cloud providers and hence the
cost for customers.  In this section, we identify that variability in reliability
exists in data centers, and study opportunities to exploit this variability to
optimize total cost of ownership (TCO).


\subsection{Asymmetric Reliability Requirements}
\label{sec:rel-variation}

We find that there are two sources of the inherent asymmetry in reliability
requirements: (1)~server/cloud applications require varying levels of reliability 
based on several factors, and (2)~there is heterogeneity in the reliability
offered by the hardware itself.


{\bf Application Resiliency Variability:} \emph{Resiliency}, or memory error tolerance,
refers to the ability of server/cloud applications to cope with memory errors.
Application resiliency can involve three important aspects: (1)~tolerating the
performance penalty of error detection or correction,
(2)~enduring potential data corruption from memory errors~\cite{luo.dsn14}, and/or
(3)~dealing with unavailability due to a server crash/reboot. Cloud applications
are known to exhibit varied resiliency to memory errors~\cite{luo.dsn14, vecc}.
We observe variation in applications' resiliency across four dimensions:
\begin{itemize}[leftmargin=*,noitemsep,topsep=0pt]

\item {\em Application role:} while certain applications, like banking and
  in-memory databases, are highly sensitive to memory errors, applications such
  as front-tier state-less applications or video streaming may be more tolerant;

\item {\em Criticality:} OS/Hypervisor regions may require high reliability,
  unlike guest virtual machines or user applications;

\item {\em Address space:} certain parts of the application address-space (e.g.,
  stack/code regions) may be more sensitive to memory errors than others (e.g.,
  heap/data regions); and

\item {\em Access mode:} read-only/clean memory areas are more amenable to
  recoverability from memory errors than written/dirty memory
  regions~\cite{luo.dsn14}.  

\end{itemize}

These dimensions of variability can
be leveraged to perform \emph{cost-effective} memory hardware provisioning ---
mapping sensitive/critical regions to reliable memory hardware (with error correction) and high-resilient regions to less protected memory
hardware (with error detection or no protection).  Note that variation of application data
resiliency over time, due to changes in workload/client behavior, may require
these regions to be remapped to the hardware.

{\bf Hardware Health Variability:} Large-scale studies have shown that DRAM
within servers exhibits significant reliability
variation~\cite{schroeder2009dram, meza2015revisiting}.  DRAM
errors have been shown to be concentrated within a small fraction of weak cells (i.e., error prone cells or slow cells),
and the behavior of errors has shown relative stability over time~\cite{sridharan2012study, lee.sigmetrics17,
chang.sigmetrics16, chang.sigmetrics17}.  As a result,
the reliability variation of DRAM can be used to perform long-term relaxation
of memory protection.  For example, healthy DRAM DIMMs may initially be
provisioned with parity protection.
As the health of the memory degrades, the
protection can be upgraded to stronger protection (e.g., SECDED).  Cloud platforms
commonly employ simple memory health/error monitoring techniques~\cite{schroeder2009dram, meza2015revisiting},
which can be leveraged to adjust the level of error protection.


\subsection{Leveraging Reliability Asymmetry for Capacity}
\label{sec:motivation:capacity}

A key consequence of the asymmetry-aware memory provisioning discussed in
Section~\ref{sec:rel-variation} is the additional memory capacity that it
offers compared to the current one-size-fits-all provisioning approach. 
As we discussed in Section~\ref{sec:background:memory}, storage for SECDED data incurs a
12.5\% overhead.  Eliminating SECDED protection frees up this 12.5\% of storage
for additional data, while performing only error detection frees up 10.7\% of 
additional memory.


Data centers can leverage this additional memory capacity to optimize
TCO in two main ways.  First, memory is often the bottleneck
resource in determining the hosting
capacity of a cloud platform~\cite{anagnostopoulou2012barely,
lim2009disaggregated, waldspurger2002memory}.  An increase in memory capacity is likely
to correspond to an increase in the number of hosted virtual machines on
a cloud, which directly contributes to cloud revenue/profit.  Second, the impact of
memory capacity to application performance is well studied in
literature~\cite{denning.csur70, levy.ieee82, megiddo.fast03, lim2009disaggregated} -- a small amount of additional capacity, when allocated
to the right application, is known to provide non-linear performance improvements.
Cloud platforms can offer opportunistic memory allocation (similar to
ballooning~\cite{waldspurger2002memory}) to applications with high memory demand,
resulting in improved application performance and customer satisfaction.

\begin{figure}[h!]
\centering
\vspace{-9pt}
\includegraphics[trim=5 95 0 0,clip,width=.9\linewidth]{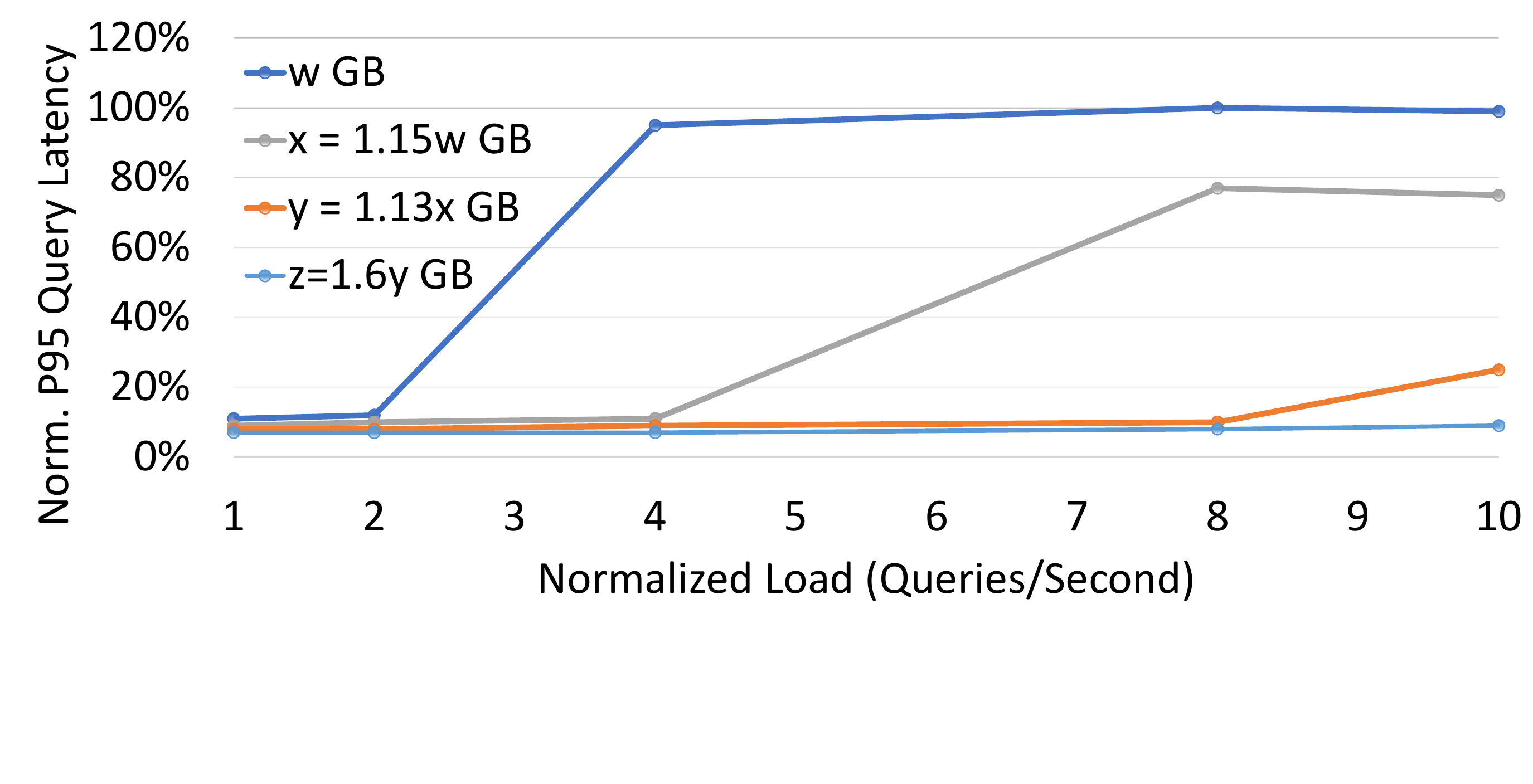}
\vspace{-6pt}
\caption{
  WebSearch 
  exhibits 37.3\% improvement on average in 95th percentile
  latency when given a 12.5\% increase in memory capacity.
}
\label{fig:websearch-capacity}
\vspace{-12pt}
\end{figure}

We quantify the performance improvement from additional memory capacity using an
interactive {\em WebSearch} cloud application from Microsoft,
running production search queries. {\em WebSearch} stores several hundred
gigabytes of search indexes in persistent storage, and uses DRAM as a cache for
storing frequently-accessed index data. We can relax the ECC protection for {\em
WebSearch} to gain a 12.5\% capacity increase, as prior work has shown that
web search applications can tolerate a large number of memory
errors~\cite{luo.dsn14}. Figure~\ref{fig:websearch-capacity} shows that memory
capacity plays a crucial role in the workload percentile latency.\footnote{For
business reasons, we anonymize the latency and capacity numbers.} We normalize
both the percentile latency on the y-axis and the load on the x-axis to their
largest observed values. Each curve shows the percentile latency for {\em
WebSearch} with different memory sizes, $w$, $x$, $y$, and $z$. By comparing
these curves, we make two observations. First, we look at these curves under the
highest normalized load (10).  We find that if error protection is eliminated, an approximately
12.5\% increase in capacity results in significant latency improvement (e.g.,
67\% from $x$ to $y$, and 24\% from $w$ to $x$). It is well known that latency
plays a crucial role to revenue of cloud workloads~\cite{hamiltonBlog,
greenberg2008cost}. Thus, it is desirable to keep the percentile latency low.
Second, we look at the highest load that guarantees a low percentile latency
(e.g., 20\% on the y-axis).  We find that, by increasing memory capacity by
about 12.5\%, load capacity for {\em WebSearch} doubles.

\subsection{Need for Dynamic DRAM Error Protection}

As we can see, there are tangible benefits to exploiting reliability variation 
in DRAM to increase its capacity.  Realizing these capacity benefits relies on
the server/cloud to offer \emph{heterogeneous} and \emph{configurable} error
protection in memory.  Though it is possible to statically provision error
protection by using different memory hardware across servers/clusters, this
approach has two key limitations: (1)~the optimal amount of memory allocated
for a certain level of protection may vary over time due to changing workload 
and hardware behavior, which could result in under- or over-provisioning when
using static partitioning; and (2)~sourcing server hardware components relies
on pricing advantages associated with procuring commodity components in bulk,
which will be disrupted if DIMMs with different reliability schemes must be 
procured.


We envision a cloud that can dynamically configure its memory resources, both 
within and across servers, to offer any combination of memory error protection
based on varying application/hardware demands.  Our goal in this work is to
design a mechanism that can dynamically repartition a single type of DRAM to
support multiple reliability schemes.

\section{CREAM Design}
\label{sec:design}

As we see in Section~\ref{sec:motivation}, there are several applications that
do not require error correction, and can benefit from additional DRAM
capacity.  However, while ECC DRAM provides additional raw capacity within each
DRAM module to store error-correcting codes, this capacity \emph{cannot} be used by
applications that do not require error correction.  In this work, we
propose \emph{Capacity- and Reliability-Adaptive Memory} (CREAM), a hardware 
mechanism that allows applications without strong reliability requirements to 
exploit the additional ECC DRAM capacity to store more user data (and reduce 
the number of page faults).


CREAM exposes the additional DRAM capacity by rearranging how data is stored
in a portion of the ECC DRAM.  In CREAM, part of the DRAM supports error detection
or no correction/detection, for applications, memory regions, or highly-reliable DRAM that do not 
require it, while part of the DRAM continues to support SECDED correction for
others that require high reliability.  The size of the two
parts can be adjusted dynamically, based on the mix of applications being run
on the server, and on the health of the DRAM.

Figure~\ref{fig:baseline} shows how data is traditionally stored alongside the 
SECDED code within an ECC DRAM.  The layout of data remains unchanged for the
high-reliability portion of DRAM in CREAM.  We propose several solutions to
to rearranging data \emph{when no correction or detection is required}
(Section~\ref{sec:design:nonecc}), each of which has distinct advantages and
overheads.  For all of these solutions, the effective DRAM capacity increases by 
12.5\% within the unprotected region.  We also propose a solution 
that supports \emph{error detection} (Section~\ref{sec:design:parity}), which 
can increase the DRAM capacity within the region by 10.7\% while
protecting against silent data corruption.

To support two regions of memory with different levels of reliability, CREAM
requires additional, low-cost hardware support
(Section~\ref{sec:design:mechanism}).  Small modifications are needed within the
memory controller to make it aware of the change in hardware layout.  Several,
but not all, of our solutions require a small bridge chip on the DRAM DIMM to
enable \emph{rank subsetting} (i.e., decoupling the chips within a rank so that
not all of them operate in lockstep) to optimize performance.  Prior work has
shown that rank subsetting can be enabled by a bridge chip at low
cost~\cite{minirank}.  On the software side, the OS page allocator must be
informed about the additional physical pages available in DRAM, and allocation
decisions must now take the reliability of a physical page and the required
reliability of applications into account; we consider such changes to be beyond
the scope of this work.  CREAM does not require any changes to the virtual
memory management within the processor, or to applications executing within
DRAM.

\subsection{Correction-Free Memory Regions}
\label{sec:design:nonecc}

In conventional ECC DRAM, even when correction is not required, each read or
write command fetches 72~bytes (64~bytes of data and 8~bytes of ECC information)
to the memory controller as before. By disabling the ECC in the memory
controller, the fetched ECC information is simply ignored. In such a scenario,
disabling ECC only brings minimal latency benefits (avoiding the short ECC
decoding latency), and does not provide any additional DRAM capacity.  In CREAM,
we instead propose to expose this capacity so that applications can use it to
store more data in DRAM. We next discuss several alternatives to organizing the
data when this capacity is exposed.


\subsubsection{Solution 1: Packed Data Layout} \label{sec:design:noecc1}


We first try a naive approach to utilizing the extra space available, which we
call the \emph{packed} data layout.  Since the newly-available capacity exists
on the DRAM chip that used to store the ECC data (Chip~8 in
Figure~\ref{fig:baseline}), our goal is to simply pack additional data pages
into this chip, keeping the layout of existing physical pages untouched.  As we
shall see, this approach requires \emph{no modifications to existing ECC DRAM}.



\begin{figure}[h!]
\centering
\vspace{-5pt}
\includegraphics[trim=10 100 10 0,clip,width=\linewidth]{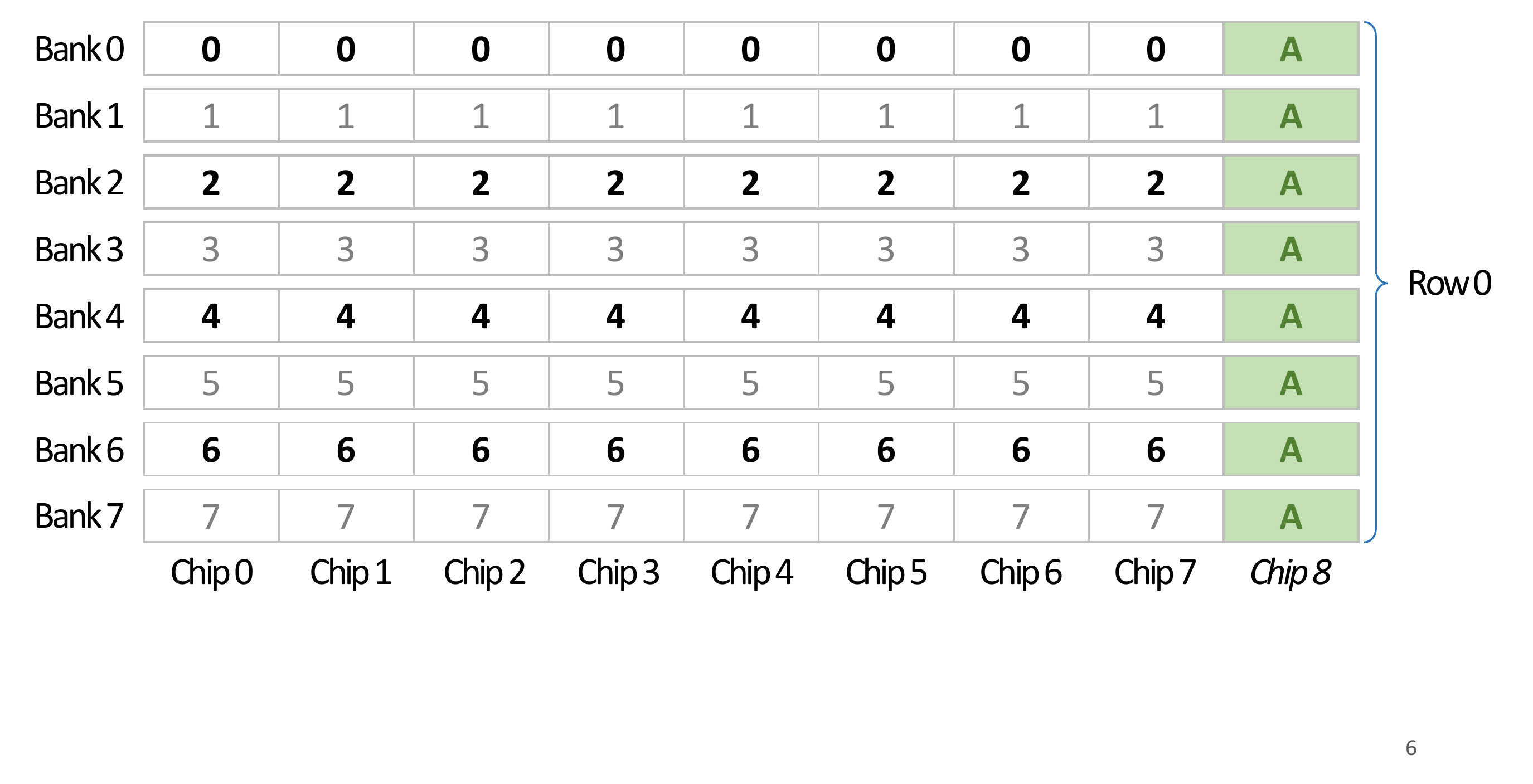}%
\vspace{-5pt}
\caption{
  Packed data layout (Solutions~1 and~2).  Page~A shows how extra
  capacity is allocated within this layout.  Compare with 
  Figure~\ref{fig:baseline}, top.
}
\label{fig:noecc1}
\vspace{-5pt}
\end{figure}

Figure~\ref{fig:noecc1} illustrates how we use the extra space. This entire
figure shows the data layout for the first DRAM row (i.e., Row~0) of each bank.
Each column of the table represents a single chip. Each entry in the table shows
the physical page number of the data stored in the corresponding chip and bank.
Note that the data layout for Pages~0--7 remains the same as the baseline
(Figure~\ref{fig:baseline}, top). As was the case before, each cache line in
these pages is striped across Chips~0--7, such that the entire page can be
stored in one row of a bank across the first eight chips (e.g., Page~0 is stored
only in Row~0 of Bank~0).  The extra page within this DRAM row, however, is only
stored within Chip~8, instead of being striped across eight chips, as Chip~8 is
the only vacant chip.  As Figure~\ref{fig:noecc1} shows, we break extra Page~A
into eight parts, and distribute each of these parts \emph{across all eight
banks}.\footnote{If we were to instead distribute the parts of one page across several
rows within a single bank, multiple accesses \emph{within a page} could incur
row buffer misses.}
Unlike Pages~0--7, where \emph{each} cache line is striped
across multiple chips, the cache lines of Page~A are instead kept within a
single bank.

{\bf Access Latency:} Recall from Section~\ref{sec:background} that a single
read operation reads data from all nine chips, retrieving 72~bytes of data over
eight data bursts.  As was the case in the baseline, when reading a cache line
from Pages~0--7, only a single read operation is required.  In this case, the
data retrieved from Chip~8 is simply ignored, as it belongs to some part of
Page~A.\footnote{It is possible to cache the data from Page~A in the 
memory controller and hope that it will be accessed in the near future.  We do 
not add such a cache, as we expect that this data, which resides in a different
OS page, is unlikely to be used within a short timespan.}  In contrast, reading
a cache line from one of the packed extra pages, such as Page~A, now requires
\emph{eight} back-to-back read operations, as each read operation only retrieves
8~bytes (i.e., 8~bits/burst) of useful data from Chip~8.  
As all of the cache line from the extra Page~A is stored within a single bank,
there continues to be at most one row miss, as once the row is activated, all
eight read operations go to different columns within the same row.  

All write operations must now be performed as read-modify-writes (i.e., data
must now be read first into the memory controller and modified there before
making changes to DRAM).  This is because writes also continue to access all 
eight chips in parallel.  For example, when we write a cache line in Page~0,
8~bytes belonging to Page~A is also overwritten.  Therefore, we must first
read the data from Page~A into the memory controller with a single read 
operation, so that we write back the same data to Page~A (thus leaving Page~A's
contents unmodified).  A write to a cache line in Page~A requires eight write
operations, for the same reason that multiple read operations were required.

{\bf Parallelism:} While the number of banks remains unchanged between the
baseline ECC DRAM and Solution~1, the degree of memory-level parallelism may 
drop slightly.  Requests to extra pages have a longer occupancy within DRAM,
reducing the overall request throughput.


In conclusion, our packed data layout exposes additional data capacity
\emph{without modifying the ECC DRAM DIMM}, but the high latency to
extra pages and for write operations may negate the effects of added
capacity.

\subsubsection{Solution 2: Rank Subsetting} \label{sec:design:noecc2}


While Solution 1 (packed data layout) enables us to utilize the ECC chip capacity,
it has two major drawbacks that may result in performance degradation and 
increased energy consumption.
First, writing data to \emph{any} page now requires a read-modify-write. 
Although writes are not usually on the critical path, the added write latency
can still delay subsequent reads that are on the critical path, and 
also increases energy consumption.  Second, an access to the extra page within
Chip~8 can disrupt the row buffer locality of accesses to a regular page within
Chips~0--7, even though the data for these two pages resides in completely
different chips.  This is a limitation of the fact that all chips within a rank
are wired to operate in lockstep.


To reduce unnecessary data transfers and reduce DRAM
energy, we employ \emph{rank subsetting}, which separates the nine chips within
a rank into two subsets, similar 
to prior work on mini-ranks~\cite{minirank}. Each rank subset can be controlled
independently, and thus can access different addresses in parallel. \emph{Within} each
subset, the chips continue to operate in lockstep.  Chips~0 to 7
form an x64 rank subset (i.e., the subset delivers 64~bits of data during each
data burst).  An x64 rank subset operates the same as a conventional
\emph{non-ECC} DIMM.  Chip~8 forms its own x8 rank subset, which has an 8-bit
bus width. An x8 rank subset still requires eight DRAM accesses (or 64
bursts) to fetch a cache line split across eight columns in a row, the same as
in Solution~1. Rank subsetting is enabled using a small bridge chip on the DRAM
DIMM, which can control chip enable signals based on which subset is currently
being accessed~\cite{minirank} (we discuss this further in
Section~\ref{sec:design:bridge}).  Note that we continue to use the data layout
from Solution~1 (Figure~\ref{fig:noecc1}).


{\bf Access Latency:} Compared to Solution~1, rank subsetting allows us to
eliminate reading from or writing to data other than the cache line being operated
on, as only the subset of
chips containing the cache line data is enabled during a memory operation. As a result, it
eliminates the need to perform read-modify-writes for every write request, as
the chips containing unmodified data are simply disabled. Note that while
this solution eliminates all redundant data transfer, each read request to the
extra page (i.e., the x8 rank subset) still requires eight accesses.

{\bf Parallelism:} Rank subsetting allows us to access
both subsets in parallel.  As the two subsets are now decoupled from each
other, a request to Chip~8 no longer disrupts the row buffer locality within
Chips~0--7.  
However, since requests to the x8 rank subset (i.e., Chip~8)
still require eight read/write operations, the bank-level parallelism is not
doubled as a result of rank subsetting.

In conclusion, adding rank subsetting to our packed data layout eliminates the
need for read-modify-writes with the assistance of a small bridge chip on the
DIMM, \emph{reducing the number of additional accesses}.
However, reads to the extra pages still incur a high latency, as they still
require eight back-to-back read operations.

\subsubsection{Solution 3: Wrap-Around Data Layout}
\label{sec:design:noecc3}


While rank subsetting in Solution~2 reduces energy consumption by eliminating
unnecessary chip accesses, accesses to Chip~8 (i.e., the x8 rank subset)
continue to require \emph{eight} DRAM operations.  Assuming that memory accesses
are uniform across all pages, the \emph{average} number of DRAM accesses
\emph{across all pages} increases by 78\%.\footnote{Smart memory allocation
could allocate \emph{cold pages} (i.e., pages with the least number of accesses)
into Chip~8, thus minimizing the total number of extra memory operations.
However, this requires software support to identify cold pages, which we do not
evaluate in this work.}

\begin{figure}[t]
\centering
\includegraphics[trim=10 100 10 0,clip,width=\linewidth]{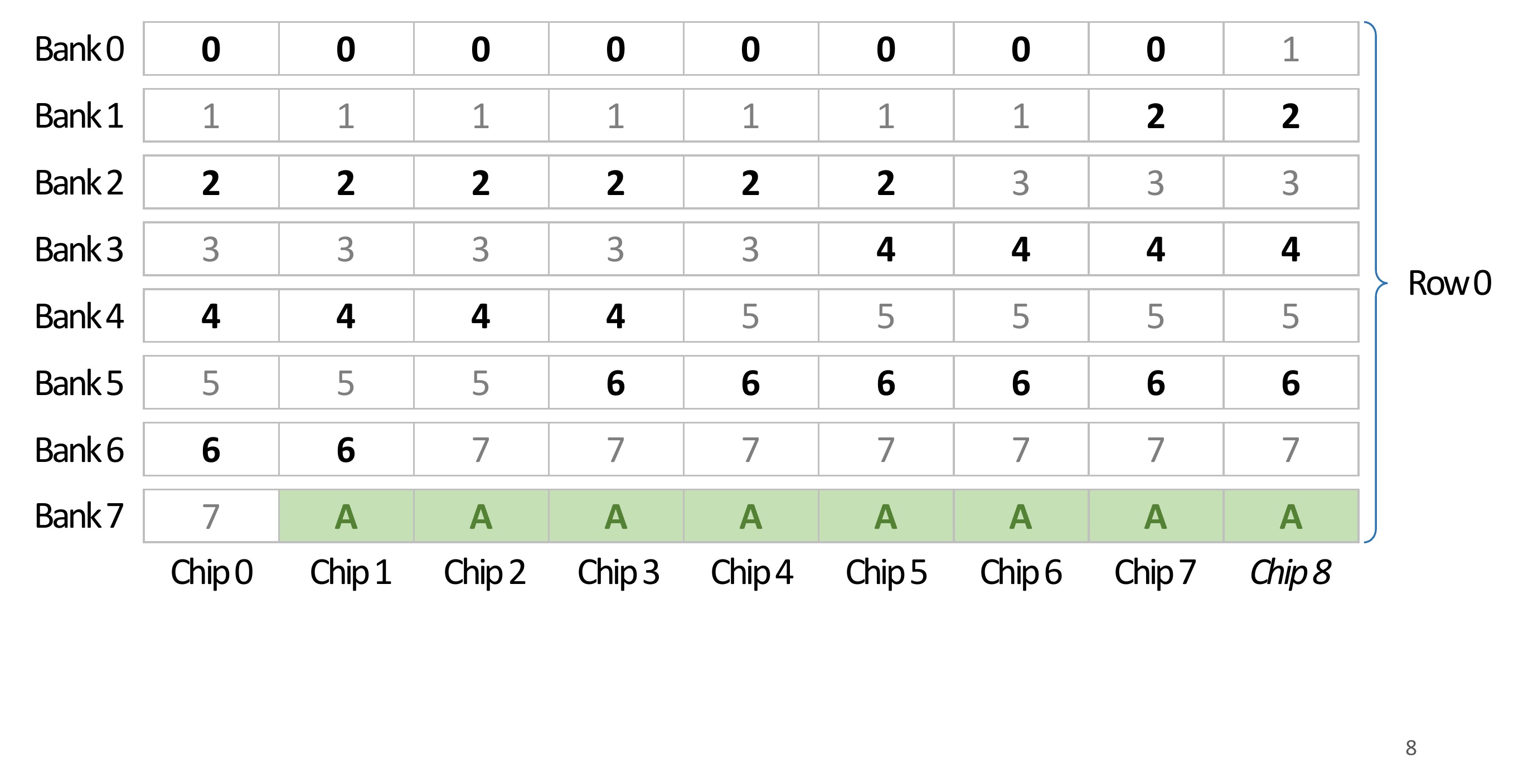}%
\vspace{-5pt}
\caption{
  Inter-bank wrap-around (Solution~3).  Page~A shows how extra
  capacity is allocated within this layout.  Compare with 
  Figure~\ref{fig:baseline}, top.
}
\label{fig:noecc3}
\vspace{-15pt}
\end{figure}

We propose a new solution, \emph{inter-bank wrap-around}, that takes advantage
of rank subsetting to ensure that \emph{every} cache line access can now be
completed in a single operation.  As each chip can still only return 8~bits in
each data burst, we must completely rearrange the data layout such that all
cache lines, including those in the extra pages, are striped across eight chips.
Figure~\ref{fig:noecc3} illustrates how we achieve such a layout, showing the
data layout for the first DRAM row of each bank (i.e., Row~0). Each row in the
figure represents a DRAM bank within the first DRAM row, and each column of the
table represents a DRAM chip.  The original mapping of pages across the first
eight rows in Bank~0 is shown in the bottom of Figure~\ref{fig:baseline} for
reference.  As is the case in the baseline, Bank~0 in our new layout contains
Page~0, except for Chip~8.  In the baseline, Page~1 mapped to Bank~1, across
Chips~0--7.  In our new layout, we move the data for Page~1 previously stored in
Bank~1, Chip~7 into Bank~0, Chip~8, causing the page to wrap around over two
banks. In this data layout, we can modify our rank subsetting logic such that
the bridge chip dynamically selects \emph{any} eight chips to be operated on at
a time.  Thus, to access Page~1, the bridge chip now opens the first row of
Bank~1 in Chips~0--6, as well as the first row of Bank~0 in Chip~8, and does not
touch Chip~7.  Likewise, as we show in Figure~\ref{fig:noecc3}, we wrap around
the remaining pages, allowing us to fit nine pages within eight banks.  In this
layout, we assign the extra Page~A to Chips~1--8 of Row~7, taking up the extra
space freed up by wrapping around the eight pages that originally resided in
these eight rows.

{\bf Access Latency:} In this data layout, \emph{all} data is striped across eight
chips. Compared to the packed data layout solutions (Solutions~1 and~2), no
cache line requires extra memory accesses, and thus memory latency is minimized.

{\bf Parallelism:} Compared to the baseline ECC DRAM, Solution~3 can in fact
improve the bank-level parallelism within a DRAM module. Thanks to rank
subsetting, \emph{each chip} can now operate in parallel.  In total, there are
8~banks $\times$ 9~chips = 72~independently operable \emph{bank slices}. For
Solution~3, each DRAM access requires eight different bank slices to supply data
at the same time to eliminate extra accesses (as we discussed in
Section~\ref{sec:design:noecc2}).  Since each DRAM row shares the same
data layout, the 72~bank slices form nine independent groups, each containing
eight bank slices that are always accessed together. Thus, we are able to
sustain \emph{nine} concurrent requests at any time, as opposed to eight in
baseline ECC DRAM. For example, the nine pages, Pages~0--7 and~A, shown in
Figure~\ref{fig:noecc3} can be accessed in parallel.\footnote{If we wrap around
multiple DRAM rows instead of DRAM banks, the 72~bank slices no longer form nine
independent groups thus cannot achieve the same parallelism as we do.}

In conclusion, inter-bank wrap-around eliminates all additional operations for
memory requests, and increases the bank-level parallelism beyond that of the
baseline ECC DRAM.  As a result, we expect that inter-bank wrap-around can
provide performance benefits over the baseline ECC DRAM \emph{beyond} the
benefits of simply providing extra DRAM capacity.

\subsection{Detection-Only Memory Regions}
\label{sec:design:parity}

So far, we have proposed solutions that do away with error protection in memory
\emph{entirely}.  However, as we discussed in Section~\ref{sec:motivation}, 
there are applications that can loosen reliability requirements somewhat, but 
are unable to tolerate \emph{silent data corruption}.  For such applications,
even if we cannot \emph{correct} the error, simply \emph{detecting} the error is
sufficient.  For an 8-bit parity code (which detects one error per data burst,
or up to eight errors per cache line), we can still provide 10.7\% greater 
effective DRAM capacity to applications.  To this end, we propose a data layout
solution for \emph{8-bit parity}.

\begin{figure}[h!]
\centering
\vspace{-5pt}
\includegraphics[trim=10 150 10 0,clip,width=\linewidth]{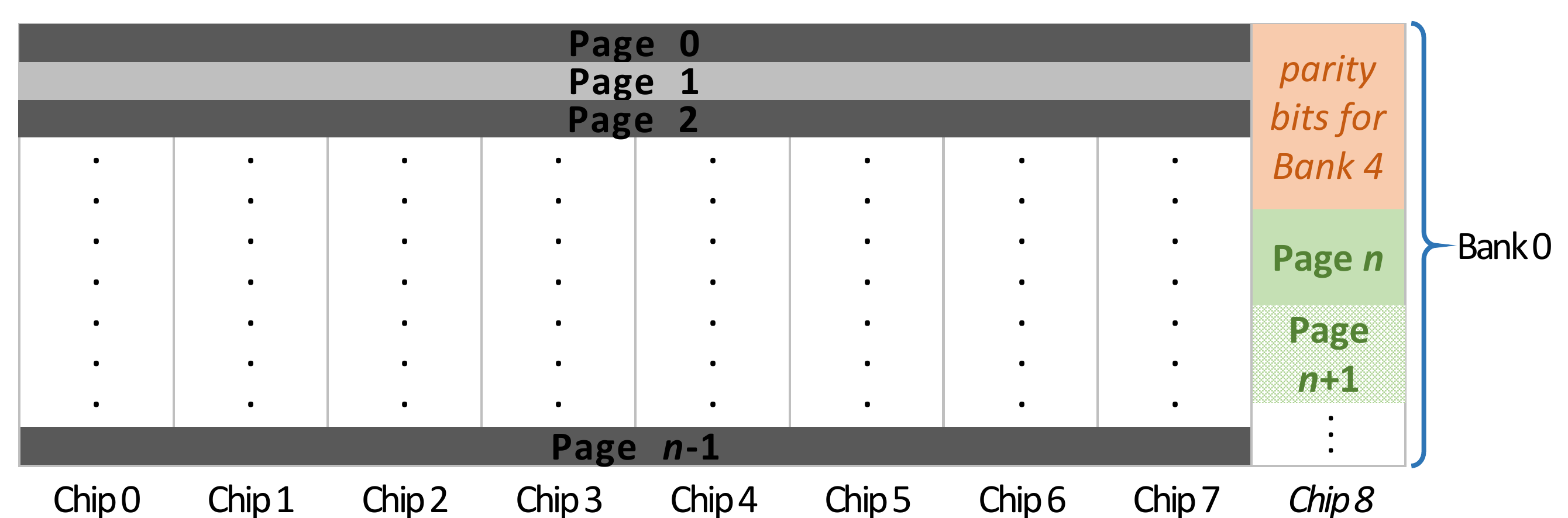}%
\vspace{-5pt}
\caption{
  Data layout \emph{within an entire bank} for 8-bit parity per cache line.  
  Pages~0 through $n$-1 each take up one row, across Chips~0--7.  Pages~$n$ and~$n$+1 
  show how extra capacity is allocated, similar to the packed data layout.
}
\label{fig:parity_layout}
\vspace{-5pt}
\end{figure}

Figure~\ref{fig:parity_layout} shows how data is laid out for 8-bit parity.
Note that this figure shows the \emph{entire} bank to simplify the explanation, 
but that the solution can also be applied to a portion of a bank.  In order to 
reduce the complexity of addressing logic, we base the 8-bit parity solution 
on the rank subsetting solution with the packed data layout
(Section~\ref{sec:design:noecc2}).  Within a bank, the physical pages that were
available already in the baseline ECC DRAM (Pages~0 through $n$-1) stay in the
same position, with each page occupying one row across Chips~0--7.  In Chip~8,
where space has been freed up from the SECDED codes, we first place parity
information.  Beyond that, the remaining free space within Chip~8 is used to
allocate extra pages, such as Page~$n$, in a packed format (i.e., the page is
split across eight rows).  As was done in Section~\ref{sec:design:noecc2}, we
employ two rank subsets: one covering Chips~0--7, and the other covering Chip~8.

{\bf Access Latency:} For read requests to the first $n$ pages, two read 
operations are performed: one for the data from Chips~0--7, and the other for 
the parity data from Chip~8.  
On a write, since the Chip~8 data contains
parity information for other cache lines, a read-modify-write
is again required to avoid modifying
the parity information for unmodified cache lines.
For extra pages, such as Page~$n$, a read request requires \emph{nine} 
operations to complete, with eight read operations to retrieve the data itself, 
and a ninth read operation to retrieve the parity code.  A write request 
requires eight write operations for the data, and a read-modify write
to save the parity data without
changing the parity information for other cache lines.  In order to avoid row
buffer conflicts when the parity information is read for the extra bits, the
parity information for Bank~$i$ is saved in Bank~($i$+4)$\mod$8, minimizing
the probability for spatial locality.

Unfortunately, since the parity data is much smaller than the data received
from a single chip during a read operation, it is difficult to avoid performing
a read-modify-write for the parity data.  Currently, each row of parity in
Chip~8 contains the parity data for eight pages.  Other data layouts, and 
perhaps layouts for other error detection encodings, can be employed to improve
performance, but we leave such studies for future work.

\subsection{Enabling Adaptive Capacity and Reliability}
\label{sec:design:mechanism}


The various solutions for CREAM require relatively simple hardware support.
Solution~1 requires modifications only within the memory controller, while
Solutions~2 and~3 add simple logic to a bridge chip on the DRAM module.  
\emph{No changes are required inside the DRAM chips.}  We
now discuss these modifications in detail, assuming an initial address space of
8GB on the ECC DRAM module to simplify our explanations.  We quantify the
overhead in Section~\ref{sec:design:overhead}.

\subsubsection{Memory Controller Support}
\label{sec:design:controller}


To support both ECC and non-ECC data on the same DRAM module, the
memory controller stores a \emph{boundary} between physical pages with
conventional layout and those with CREAM layout in a register. This boundary
can be used to determine the size of the total physical address space, since it
tells us how much extra memory is added from the non-ECC portion.  For an 8GB memory,
this is 
8GB$+(\textit{boundary}\gg3)$. The physical pages within the boundary use the CREAM
data layout and store non-ECC data. The pages mapped to Chips 0--7 in the CREAM layout
(e.g., Pages~0--7 in Figures~\ref{fig:noecc1} and~\ref{fig:noecc3}) are mapped
to physical addresses from 0 to \emph{boundary}. The extra pages
(e.g., page~A) are mapped to physical addresses ranging from 8GB to the end.
Physical pages outside of the boundary use the conventional layout and store ECC
data. These pages are mapped to physical addresses between \emph{boundary} and
8GB. The simple boundary has two benefits: (1)~only the address is necessary
to identify whether a page has error correction; and (2)~as non-ECC pages are
arranged at the beginning of the physical address space, the address offset of
the extra pages is easy to calculate.
Note that
for Solutions~2 and~3, the memory controller needs to communicate this boundary
with the bridge chip, where the address translation takes place.

For Solution~1, all of the logic for CREAM, including address translation logic,
is implemented within the memory controller, so the ECC DRAM modules \emph{do
not require any modification}. The memory controller translates each read request
to the extra pages into eight back-to-back cache line accesses.  The eight
accessed addresses, \emph{ACC}, can be easily obtained from the requested
address, \emph{REQ}: $\textit{ACC} = (\textit{REQ}-$8GB$)\ll3 + 0/1/.../7$. To
assemble the requested cache line, the memory controller buffers and combines
the partial data from Chip~8 of these eight accessed cache lines within a 64B
shift register that we add to the memory controller. The same shift register is
reused to stage data during the read-modify-write operation for all pages.
These modifications are unnecessary for Solutions~2 and~3.

\subsubsection{DRAM Module Bridge Chip}
\label{sec:design:bridge}

Today's servers typically use registered memory (RDIMMs), which contain a bridge
chip on the DRAM module with logic to buffer the control and addressing
information from the memory controller.  We propose to add simple circuitry to
this existing bridge chip, to support rank subsetting and handle the proposed
address translation schemes in hardware. 


To translate the physical address of each incoming request, the bridge chip
takes the requested address sent by the memory controller, and converts it into
the rank subset enable signal for each chip and the row address for each rank
subset. Thanks to the way that we map the extra pages, when accessing any
ECC-protected data, no address translation is required. For Solution~2, the nine
chips are statically divided into two rank subsets, and the most significant bit
of the requested address determines which subset is activated.  Then, the bridge
chip, instead of the memory controller, translates the address using the same
simple logic as Solution~1.  

For Solution~3, we form two rank subsets dynamically using eight out of the
nine chips, with each subset accessing a different row within the chip. We can
determine which eight chips should be used based on the \emph{original} bank
number (i.e., the three least significant bits of the row number): the ID of the chip to be ignored is (${8 - \textit{BANK\_ID}}$).

\subsection{Hardware Overhead}
\label{sec:design:overhead}

To determine the overhead of our hardware modifications, we synthesized our 
modifications using Synopsys Design Compiler~\cite{synopsysdc},
with an open-source 14nm CMOS cell library~\cite{usc14nm}.
We find that the hardware overhead for our various CREAM solutions are very
modest.

For Solution~1, we evaluated the overhead of the address translation logic that
must be implemented within the memory controller. As a baseline, we used the
Verilog design of an FR-FCFS memory scheduler~\cite{frfcfs}.  The
modifications for CREAM increase the area overhead of the memory controller
logic by only 2.0\%.  As a comparison point, the \emph{total} memory controller
logic area comprises only 2.7\% of the area of an ARM Cortex-A72 core~\cite{cortexa72}.
We also need a 64B register to stage partial cache lines during the read and
read-modify-write operations.
We find that the logic latency of the memory controller increases by 6.3\% over FR-FCFS.
Compared to many previously-proposed schedulers, the FR-FCFS memory scheduler 
has a much lower latency~\cite{subramanian2014blacklisting}, thus the
CREAM Solution~1 scheduler should also be much faster than these other schedulers.

For Solution~3, we evaluate the overhead of the logic that we add to the bridge
chip. We find that the total area of the additional logic is only
493$\mu$m$^2$, representing less than 0.043\% of the total area of an ARM
Cortex-A72~\cite{cortexa72}. The estimated latency of the circuit is 198ps, which is much lower
than the 1 DRAM cycle latency (1.5ns in our simulations) that we conservatively
use for the bridge chip delay. We need to add 9 chip-select pins and 24 address
pins (8 sets of 3 bits, for the LSBs of the different row IDs) to the bridge chip.

\section{Methodology}
\label{sec:methodology}

{\bf Simulation Framework:} To quantitatively analyze the performance of CREAM,
we implement all three of our protection-free solutions, as well as our
detection-only solution, in Ramulator~\cite{ramulator}, a detailed DRAM
simulator. We modify the simulator to accurately model rank subsetting, and we
add a one-cycle delay for the simple translation logic (as described in
Section~\ref{sec:design:mechanism}) within the bridge chip.  The parameters of
the simulated system are summarized in Table~\ref{tbl:simulation}.  In our
simulations, we emulate the page replacement policy using an active list and an
inactive list, similar to that used in a modern Linux virtual memory
manager~\cite{gorman.book04}. We set the page fault penalty to $500\mu s$, which
includes a $300\mu s$ SSD access latency and a $200\mu s$ software
latency~\cite{saxena.hotos09}. 



\begin{table}[h!]
\centering
\scriptsize
\vspace{-5pt}
\begin{tabular}{ll}
  \toprule
    Processor & 4 cores, 2.6GHz, 4-wide issue, 128-entry ROB \\ \midrule
    Cache     & 32KB L1 cache, 512KB L2 cache, 8MB L3 cache \\ \midrule
    \multirow{2}{*}{DRAM} & 8GB DDR3-1333H, 1 channel, 1 rank, 8 banks \\ \cmidrule{2-2}
                          & Open row policy, FR-FCFS scheduler~\cite{frfcfs}\\
  \bottomrule%
\end{tabular}
\vspace{-5pt}
\caption{
  Main parameters of the simulated system.
}
\label{tbl:simulation}
\vspace{-8pt}
\end{table}

{\bf Workloads:} 
We evaluate two types of workloads: data-intensive workloads that are sensitive 
to memory capacity, and latency-sensitive workloads.

For our capacity-sensitive workloads, in addition to the WebSearch workload studied in Section~\ref{sec:motivation:capacity},
we evaluate two \texttt{memcached} configurations~\cite{memcached}. We run a synthetic
client workload that queries \texttt{memcached} for a 20GB dataset at a rate of
2430 queries/second, with the server running four threads.
The first workload configuration prevents paging, by setting \texttt{memcached}'s memory usage to 8GB and
pinning all of its resident memory in DRAM. 
The second workload configuration thrashes the physical address space across all our evaluation configurations by setting the memory usage
to 10GB. In this configuration, the \texttt{memcached} server uses more memory
space than available on the system, even when CREAM is used, and always triggers page faults. 

For our memory latency-sensitive workloads, we construct 40 multiprogrammed
four-core workloads, using applications from SPEC CPU2006~\cite{spec} and
TPC~\cite{tpcc,tpch}. We classify each application based on its number of
last-level cache misses per thousand instructions (MPKI), as has been done in
prior work (e.g.,~\cite{subramanian2014blacklisting}).  Applications with an
MPKI greater than~10 are classified as \emph{memory-intensive}, and all other
applications are classified as non-memory-intensive.  We sweep over the fraction
of memory-intensive applications within each workload, ranging from 0\% to
100\%.  For each category in the sweep, we build eight workloads by randomly
selecting memory-intensive and non-memory-intensive workloads.  Each application
in the workload is run until the \emph{slowest} application completes
200~million instructions, to ensure that realistic contention is simulated.
We quantify multiprogrammed workload performance using \emph{weighted 
speedup}, a commonly-used metric to express multicore workload
performance~\cite{snavely2000symbiotic, eyerman2008metrics}.  Weighted speedup
is calculated as the sum of speedups for each application (vs.\ a baseline
where each application runs without interference).



\section{Evaluation}
\label{sec:evaluation}

We now evaluate the performance of CREAM, our proposed mechanism to expose the
additional capacity of ECC DRAM when applications don't require strong 
reliability.  
We examine seven configurations:
\begin{itemize}[leftmargin=*,noitemsep,topsep=0pt]

  \item \texttt{Baseline:} an unmodified ECC DRAM;

  \item \texttt{Packed:} a CREAM configuration that uses the packed data layout
    (Section~\ref{sec:design:noecc1});

  \item \texttt{Packed+RS:} a CREAM configuration that uses the packed data 
    layout in conjunction with rank subsetting 
    (Section~\ref{sec:design:noecc2});


  \item \texttt{Inter-Wrap:} a CREAM configuration that uses the inter-bank 
    wrap-around data layout in conjunction with rank subsetting
    (Section~\ref{sec:design:noecc3});

  \item \texttt{Parity:} our detection-only CREAM
configuration with 8-bit parity (Section~\ref{sec:design:parity}); and

  \item \texttt{SoftECC:} a mechanism based on Virtualized ECC~\cite{vecc} that provides error correction
in non-ECC DRAM by storing SECDED information within some of the physical pages
within DRAM, \emph{lowering} the effective capacity of the DRAM by up to 11.1\%.

\end{itemize}

\subsection{Capacity-Sensitive Workloads}
\label{sec:evaluation:memcached}

We evaluate the data-intensive \texttt{memcached} workloads described in
Section~\ref{sec:methodology}.
\texttt{memcached} is typically used as a memory caching layer, which aims to
reduce the query traffic to the back-end storage layer~\cite{nishtala.nsdi13}.
However, while increasing the memory capacity of a \texttt{memcached} server can increase
its hit rate in the memory caching layer, and thus reduce the overall percentile
latency, we do not evaluate this benefit. 



\begin{figure}[h!]
\centering
\vspace{-5pt}
\includegraphics[trim=5 175 0 0,clip,width=\linewidth]{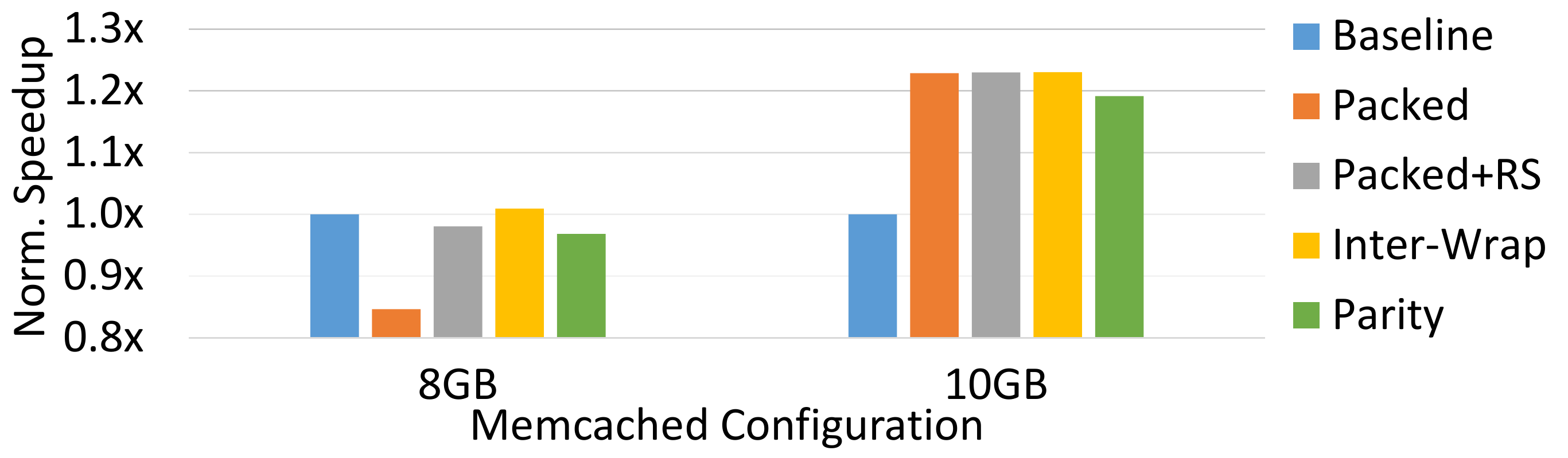}%
\vspace{-5pt}
\caption{
  \hspace{-.8em}\texttt{memcached} speedups normalized to \texttt{Baseline}.
}
\label{fig:memcached}
\vspace{-10pt}
\end{figure}

Figure~\ref{fig:memcached} plots the speedup for each \texttt{memcached}
workload. We first look at the 8GB workload
configuration, where no page faults occur in any of the systems that we
evaluate. We use this to observe the overhead of CREAM for a data-intensive
application. We find that while the overhead for \texttt{Packed} is moderate over \texttt{Baseline},
at 17.0\%, \texttt{Inter-Warp} in fact achieves a slight performance 
\emph{improvement} (of 0.8\%), as its increased bank-level parallelism outweighs
the additional latencies.  
With no effective overheads, we believe that the
WebSearch workload used in our motivational studies
(Section~\ref{sec:motivation:capacity}) will come close to the average
performance of 37.0\% reported in those overhead-free studies.

In order to understand the aggregate impact of CREAM, combining capacity
benefits and all CREAM overheads, we study the 10GB workload configuration for
\texttt{memcached}, which generates page faults under both \texttt{Baseline} and
CREAM.  This workload represents the usage scenario where page faults are
already unavoidable in \texttt{Baseline}, which can happen due to memory
ballooning~\cite{waldspurger2002memory} or application behavior.
As we see in Figure~\ref{fig:memcached}, all of the CREAM configurations show
large benefits from the added capacity, even when factoring in all overheads.
We observe that even for \texttt{Packed}, which has a high overhead in CREAM,
the added memory capacity and reduction in page faults easily overcomes this overhead.
The best CREAM configuration, \texttt{Inter-Warp}, achieves a speedup of 23.0\%.
\texttt{Parity}, our detection-only CREAM configuration, also sees
reasonable speedups of 19.1\%, though this is lower than the protection-free
configurations due to its smaller increase in DRAM capacity over \texttt{Baseline}.

We conclude that CREAM is effective at delivering significant performance 
increases for capacity-sensitive applications that do not need ECC protection.

\subsection{Latency-Sensitive Workloads}
\label{sec:evaluation:noecc}

\label{sec:evaluation:ipc}

We now evaluate CREAM on our multiprogrammed latency-sensitive workloads.
Unlike \texttt{memcached}, many applications
cannot be configured to take advantage of the increased memory capacity, but
can still benefit from the increased
bank-level parallelism provided by CREAM. For these results, we assume that
CREAM has removed all error protection from the DRAM for the CREAM
configurations, exposing an additional 12.5\% memory capacity. However, 
\emph{no capacity-related benefits are shown in these results}, as the
workloads are not sensitive to memory capacity.



\begin{figure}[h!]
\centering
\vspace{-5pt}
\includegraphics[trim=0 90 0 0,clip,width=0.9\linewidth]{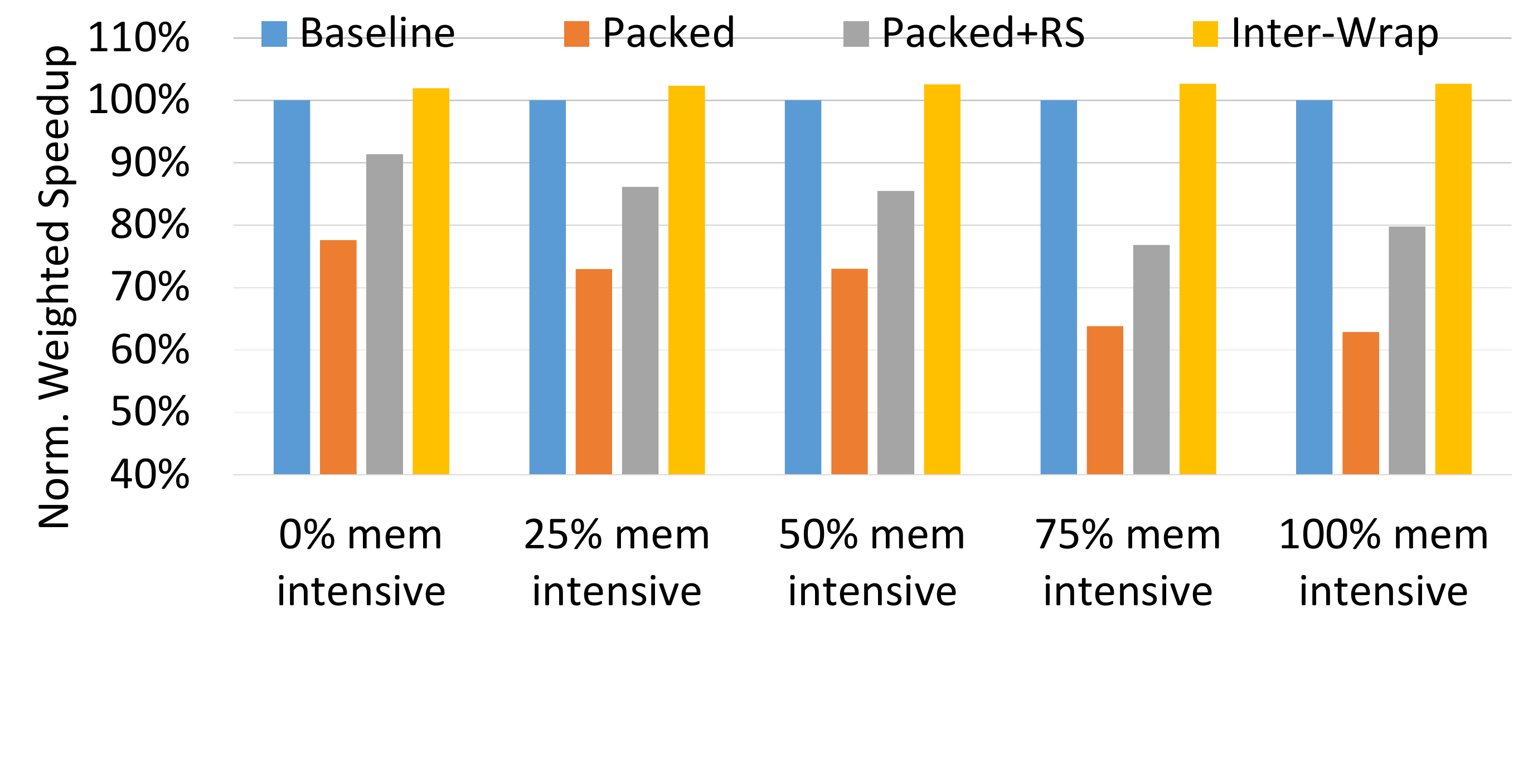}%
\vspace{-5pt}
\caption{
  Weighted speedup under different memory intensity levels, normalized to
\texttt{Baseline}.
}
\label{fig:ipc}
\vspace{-8pt}
\end{figure}

Figure~\ref{fig:ipc} shows the weighted speedup for \texttt{Baseline} and
our three CREAM correction-free configurations when the whole DRAM module
has no error correction, normalized to the \texttt{Baseline}
weighted speedup on the y-axis. On the x-axis, each group of bars represents a 
different number of memory-intensive applications within the workload (see
Section~\ref{sec:methodology}).  We make four observations from
these results:  (1)~\texttt{Packed} experiences an average performance degradation
of 29.9\%; (2)~\texttt{Packed+RS} does better than \texttt{Packed}, but still 
has an average performance degradation of 16.1\%; (3)~both \texttt{Packed} and
\texttt{Packed+RS} experience worse performance degradation as the workload
memory intensity increases; and (4)~\texttt{Inter-Wrap} \emph{improves} system
performance by 2.4\%, with greater improvements at higher memory intensities.  
We now examine why we observe these performance trends.


\label{sec:evaluation:requests}

\begin{figure}[h!]
\centering
\vspace{-10pt}
\includegraphics[trim=0 140 0 0,clip,width=0.95\linewidth]{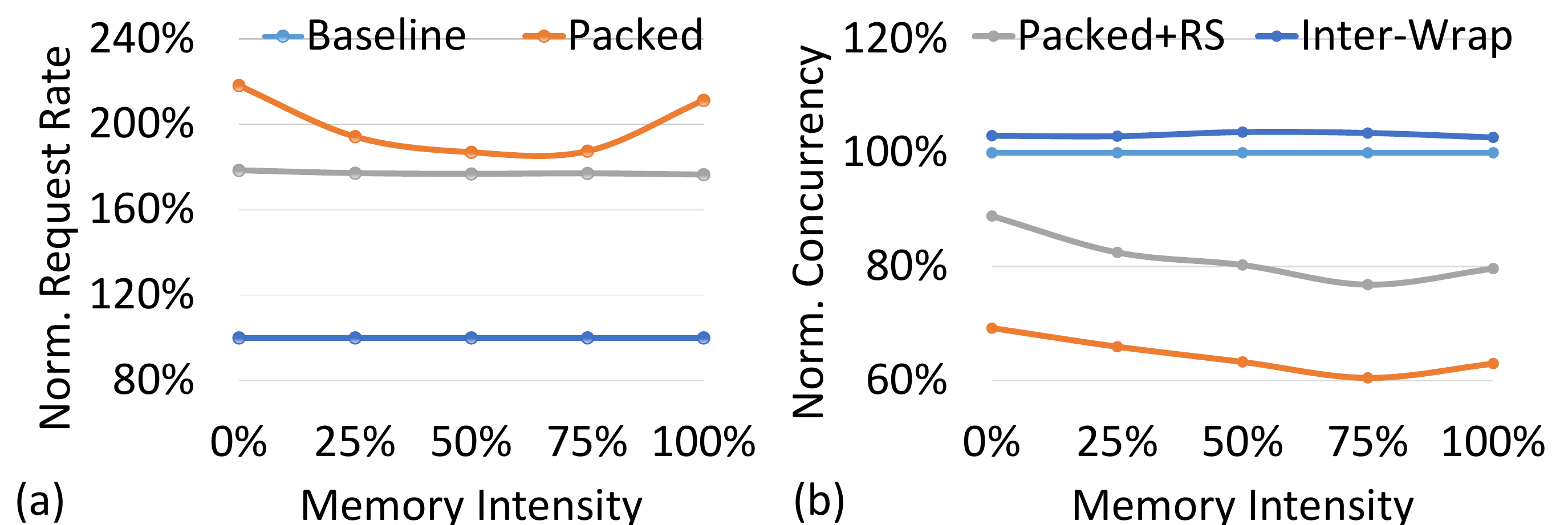}%
\vspace{-5pt}
\caption{
  (a) Number of memory requests issued for each configuration, and (b) average
number of concurrent memory requests for each configuration, both normalized to
\texttt{Baseline}.
}
\label{fig:requests}
\label{fig:parallelism}
\vspace{-5pt}
\end{figure}

\noindent \textit{\textbf{Extra Memory Requests:}}
Figure~\ref{fig:requests}a shows the number of memory requests issued by the
DRAM, normalized to \texttt{Baseline}, along the y-axis. The x-axis is the same
as in Figure~\ref{fig:ipc}. We make three observations from these results:
(1)~\texttt{Packed} effectively doubles the number of memory requests performed
on average over \texttt{Baseline}, as a result of its additional read operations
and its need for read-modify-write operations; (2)~\texttt{Packed+RS} reduces
the percentage of extra requests to an average of 77.2\% across all workloads,
which corresponds to the elimination of the read-modify-write operations that
take place in \texttt{Packed}; and (3)~\texttt{Inter-Wrap} eliminates all extra
memory requests. This agrees with our expectation from
Sections~\ref{sec:design:noecc3}, as \texttt{Inter-Wrap} rearranges all of the
pages to span across eight DRAM chips.

\label{sec:evaluation:parallelism}

\smallskip\noindent \textit{\textbf{In-DRAM Parallelism:}}
Figure~\ref{fig:parallelism}b plots the average number of concurrent memory
requests normalized to \texttt{Baseline}, shown along the y-axis. The x-axis is
the same as in Figure~\ref{fig:ipc}.
We find that this figure shows similar trends to Figure~\ref{fig:ipc}. 
This indicates that in-DRAM parallelism is a major
contributor to the performance variation across CREAM configurations.
\texttt{Packed+RS} has reduced parallelism because each memory request to data
in Chip~8 expands to eight commands, preventing other requests to the same bank from being
serviced. \texttt{Packed} reduces parallelism even more,
as the read-modify-write operations also require multiple commands per request.
In contrast, \texttt{Inter-Wrap} improves parallelism by
3.1\% over \texttt{Baseline}, because it fully utilizes all of the independent
units on the ECC DRAM to increase the effective amount of bank-level
parallelism.


\label{sec:evaluation:rowbuffer}

\begin{figure}[h!]
\centering
\vspace{-5pt}
\includegraphics[trim=0 100 0 5,clip,width=0.95\linewidth]{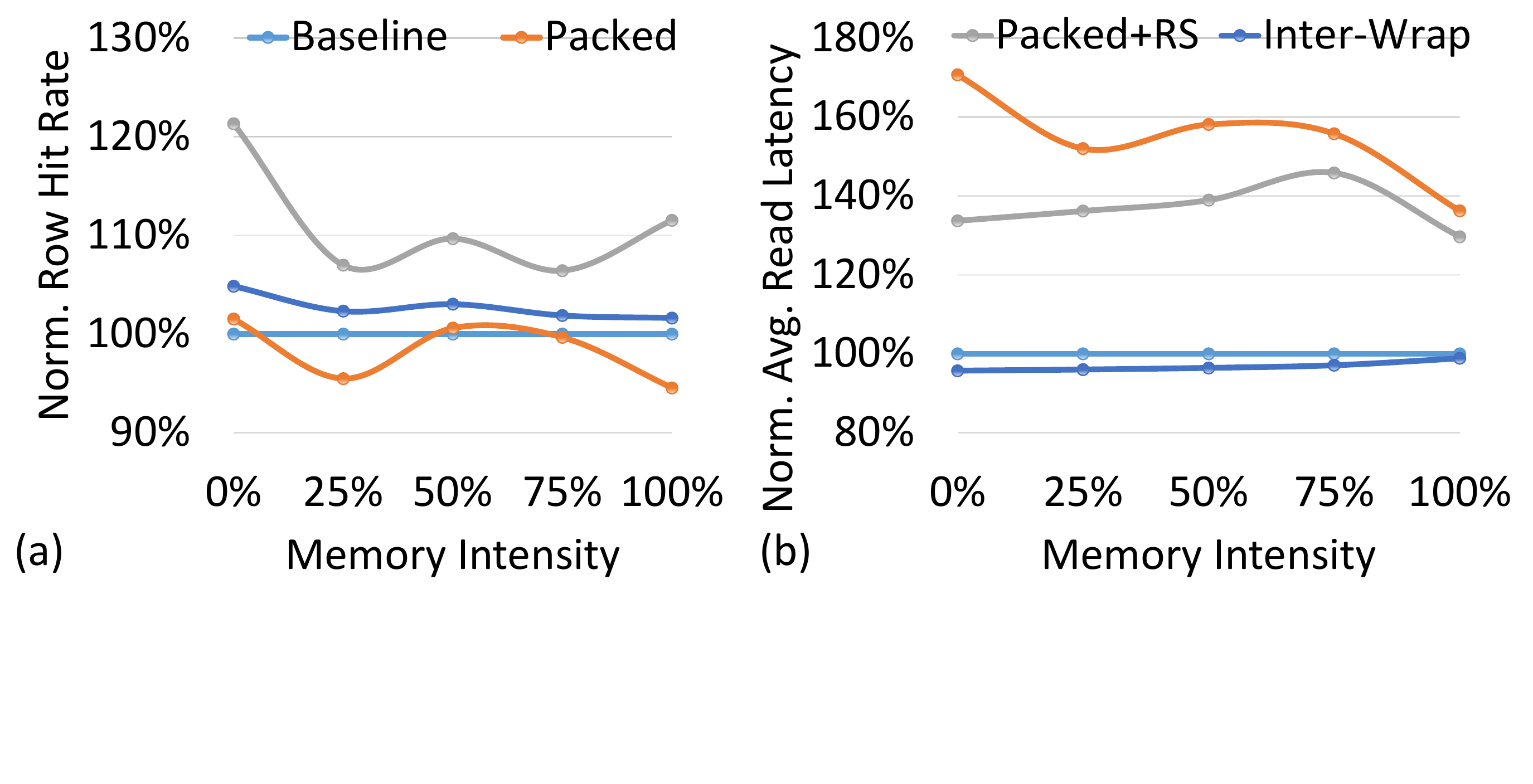}%
\vspace{-5pt}
\caption{
  (a) Row buffer hit rate, and (b) average memory latency, both
normalized to \texttt{Baseline}.
}
\label{fig:rowbuffer}
\label{fig:latency}
\vspace{-10pt}
\end{figure}

\noindent \textit{\textbf{Row Buffer Locality:}}
Figure~\ref{fig:rowbuffer}a plots the row buffer hit rate normalized to
\texttt{Baseline}, shown along the y-axis. The x-axis is again the same
as in Figure~\ref{fig:ipc}. We make three
observations from these results: (1)~\texttt{Packed} reduces the row buffer hit rate
by 1.6\%, as without rank subsetting, the number of row buffer misses increases,
but the eight commands for every request to Chip~8 counteract this by introducing
more row buffer hits; (2)~\texttt{Packed+RS} improves the
row buffer hit rate significantly, as rank subsetting eliminates the increase
in row buffer misses from \texttt{Packed}, but retains the increase in row buffer
hits due to Chip~8 requests; and (3)~\texttt{Inter-Wrap} increases the row
buffer hit rate by 2.7\%, due to
its increased in-DRAM parallelism.  Overall, we find that row buffer locality
has little impact on performance.

\label{sec:evaluation:latency}

\smallskip\noindent \textit{\textbf{Average Memory Latency:}}
Figure~\ref{fig:latency}b plots the average memory latency normalized to
\texttt{Baseline}, shown along the y-axis. The x-axis remains the same
as in Figure~\ref{fig:ipc}. We find that average memory latency
is inversely correlated with the performance, and
thus is also a major contributor to the variation across CREAM configurations.
Unsurprisingly, \texttt{Packed} and \texttt{Packed+RS} experience high average
latencies, as the additional commands per request can delay other pending memory
requests.  In contrast, the additional parallelism offered by \texttt{Inter-Wrap}
reduces memory contention, translating into shorter request latencies.



\subsection{Sensitivity Study: Correction-Free Size}
\label{sec:evaluation:sensitivity}

So far, we have assumed that the entire physical memory address space of an ECC
DRAM is transformed into correction-free memory.  In this section, we study how
the performance of CREAM changes as larger portions of the DRAM are set aside
for strong error correction (i.e., SECDED).  We compare the performance of our
CREAM configurations to \texttt{SoftECC}.  CREAM incurs no 
performance penalty for SECDED as detection and correction are already
implemented within the memory controller.  In contrast, \texttt{SoftECC} requires
modifications to the processor's Memory Management Unit (MMU) so it can issue
separate memory requests to the SECDED data, and it also utilized space in the
last-level cache to store recently-used SECDED data~\cite{vecc}.

\begin{figure}[h!]
\centering
\vspace{-5pt}
\includegraphics[width=0.9\linewidth]{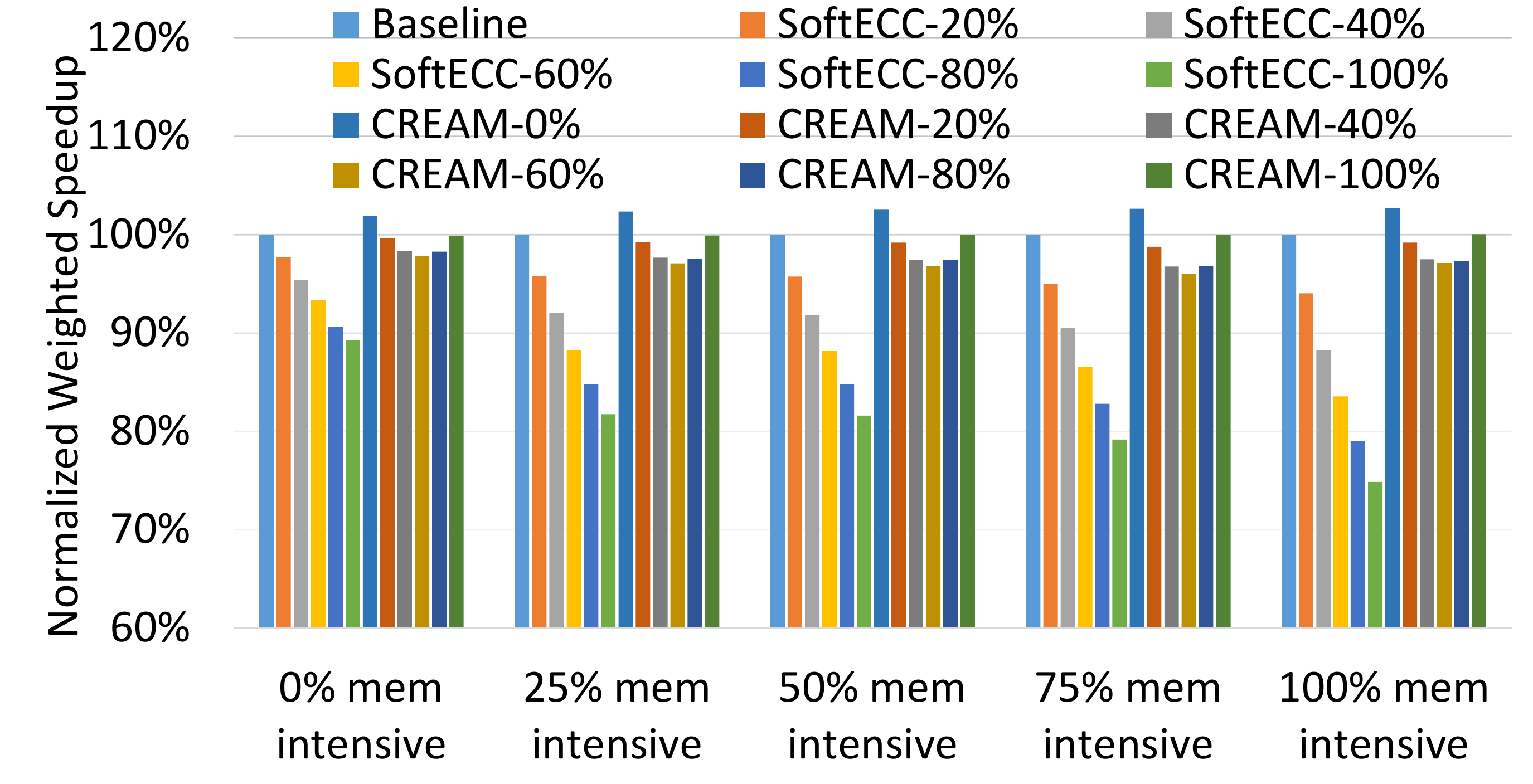}%
\vspace{-10pt}%
\caption{
  Sensitivity study on performance of \texttt{Inter-Wrap} (CREAM) and
  \texttt{SoftECC} across the fraction of DRAM allocated for SECDED correction,
  normalized to \texttt{Baseline}.
}
\label{fig:eccratio}
\vspace{-10pt}
\end{figure}

We sweep over the percentage of DRAM reserved for SECDED correction.
Figure~\ref{fig:eccratio} plots the weighted speedup, normalized to
\texttt{Baseline}, along the y-axis. The first six bars in each group
show the performance of the \texttt{SoftECC}
configuration (as no error correction is required, \texttt{Baseline} is the same
as \texttt{SoftECC-0\%}). The remaining six bars show the performance of
\texttt{Inter-Wrap}, the best of our CREAM solutions.  

We make three key
observations from this data: (1)~as the memory intensity of the workload
increases, the performance of \texttt{SoftECC} decreases, which occurs because 
\texttt{SoftECC} uses last-level cache space to store ECC data, increasing the
cache contention; (2)~as the percentage of DRAM using SECDED increases,
\texttt{SoftECC} performance also drops, as much as 25.1\% at our highest 
memory intensity; and (3)~across all proportions of SECDED-covered DRAM, 
\texttt{CREAM} maintains minimal performance degradation, with the lowest
performance drop being only 4.0\%.  The small performance drops for CREAM
occur when there is a balance between the amount of SECDED-covered DRAM and
correction-free DRAM (the worst performance occurs at 60\% SECDED coverage),
because a SECDED-covered cache line destroys row buffer locality for up to two
rank subsets that were being used by a correction-free cache line.  

We conclude that these impacts are minimal, and that even setting aside the performance
improvements from CREAM's larger memory capacity, CREAM delivers very low
performance impacts when switching between SECDED-covered and correction-free DRAM
regions across the entire range of our sensitivity study.

%
%


\section{Related Work} \label{sec:related}

To our knowledge, this paper is the first to (1)~exploit the ECC storage within
an ECC DRAM module as extra memory capacity for applications or memory regions that do not require high
reliability, and (2)~propose a hardware mechanism to rearrange the data layout in an
ECC DRAM module to efficiently exploit the extra memory capacity.

We have already compared the performance of our work, CREAM, to
a mechanism similar to Virtualized ECC~\cite{vecc} in Section~\ref{sec:evaluation:sensitivity}.
Virtualized ECC (VECC) uses software to map ECC bits onto non-ECC DRAM modules, providing
flexibility between the reliability and capacity provisioned in the memory.
We show that VECC can adversely impact performance in
some cases, whereas CREAM is much more graceful: the worst-case
performance degradation of VECC over using a baseline ECC DRAM module is
25.1\%, while CREAM's is less than 4\%. In addition, CREAM provides 12.5\% extra data capacity in
the DRAM module when ECC protection is not required, while
Virtualized ECC \emph{reduces} data capacity by 11.1\% when ECC protection is
used for all data. Virtualized ECC requires hardware changes to the MMU, as well
as OS support to allocate physical pages for ECC storage. 
CREAM requires hardware changes to only the memory controller and the
bridge chip on the DRAM module, and does not require OS support (as it is handled in
hardware).

There has been a lot of work on providing flexible, efficient, and more powerful
ECC protection in DRAM~\cite{vecc, yoon.isca12.boom,
yoon.isca12.dgms, udipi2012lot,
kim2015bamboo, jian2013adaptive, li2012mage, gongclean}, as well as flexible latencies or supply voltage in DRAM~\cite{sridharan2012study, lee.sigmetrics17,
chang.sigmetrics16, chang.sigmetrics17}.  None of these works make use
of the space reserved for ECC to gain
higher capacity. Prior work has proposed in-DRAM ECC
correction mechanisms~\cite{kang2014co} (as opposed to correction in the controller). CREAM can potentially be extended for such
devices with in-DRAM ECC mechanisms, to exploit the
extra capacity dedicated for ECC when the reliability guarantees provided by ECC are
not required. 


%

Many prior works have proposed to change the data layout~\cite{yoon.isca12.boom,
yoon.isca12.dgms} or use rank subsetting~\cite{minirank, ahn.cal09} on an
ECC or non-ECC DRAM module for various reasons. None of these works use either technique to
efficiently gain data capacity from the space reserved on an ECC DRAM module for correction codes.


\section{Conclusion}
\label{sec:conclusion}

ECC DRAM, widely used in today's large-scale server systems, adds an extra DRAM
chip to each DRAM module to store error-correcting codes required for
increased reliability.  While some applications or memory regions require the error 
protection offered in ECC DRAM, others do not need
error correction.  Even though these other applications or memory regions
may benefit from additional DRAM data capacity, the extra capacity within
ECC DRAM is not available for them, as it is exclusively used for strong error 
protection codes.

In this work, we propose \emph{Capacity- and 
Reliability-Adaptive Memory} (CREAM), a mechanism that exposes the additional 
ECC DRAM capacity to those applications that do not require error 
correction.  CREAM converts a part of the ECC DRAM space to provide either no
correction or lightweight error detection, freeing up space previously used by
error-correcting codes for use as additional data capacity within DRAM.
We perform experiments with two large-memory workloads, and find that
the additional data capacity that CREAM can deliver improves their performance 
significantly.
We find that CREAM can deliver this additional data capacity without any
significant performance overhead.  We conclude that CREAM is a practical
mechanism that enables the use of capacity that is otherwise used for error
correction in modern ECC DRAM modules for data storage, thereby leading to
significant performance improvements and a new capability to efficiently trade
off between reliability and memory capacity.




\bibliographystyle{ieeetr}
\bibliography{IEEEabrv,refs}

\end{document}